\documentclass[opre,sglanonrev]{informs4_arXiv}
\usepackage{eqndefns-left} 
\RequirePackage{tgtermes}
\RequirePackage{newtxtext}
\RequirePackage{newtxmath}
\RequirePackage{bm}
\RequirePackage{endnotes}

\OneAndAHalfSpacedXII 

\usepackage{amsmath}
\usepackage{amsfonts}
\usepackage{amssymb}

\usepackage{booktabs}
\usepackage[nocomma, short]{optidef}
\usepackage[dvipsnames]{xcolor}
\usepackage{hyperref}
\hypersetup{
    colorlinks=true,       
    citecolor=black,        
    linkcolor=black,        
    urlcolor=black          
}

\newcommand{\dd}{\mathrm{d}}
\newcommand{\E}{\mathbb{E}}

\usepackage{natbib}
 \bibpunct[, ]{(}{)}{,}{a}{}{,}%
 \def\bibfont{\small}%
\EquationsNumberedThrough    

\TheoremsNumberedThrough     
\ECRepeatTheorems  %

\MANUSCRIPTNO{MOOR-0001-2024.00}

\begin{document}


\RUNAUTHOR{Zhikang et al.}

\RUNTITLE{Optimal Mediation Mechanism in Bilateral Trade}

\TITLE{Optimal Mediation Mechanism in Bilateral Trade}
\ARTICLEAUTHORS{%
\AUTHOR{Zhikang Fan}
\AFF{School of Business and Management, Hong Kong University of Science and Technology, \EMAIL{zhikangfan@ust.hk}}

\AUTHOR{Weiran Shen}
\AFF{Gaoling School of Artificial Intelligence, Renmin University of China, \EMAIL{shenweiran@ruc.edu.cn}}

\AUTHOR{Shaojie Tang}
\AFF{School of Management, University at Buffalo, \EMAIL{shaojiet@buffalo.edu}}

\AUTHOR{Yao Wang}
\AFF{School of Management, Xi’an Jiaotong University, \EMAIL{yao.s.wang@gmail.com}}

} 

\ABSTRACT{%
We study the problem of designing revenue-maximizing mechanisms for a selfish mediator who facilitates trade between a buyer and a seller. We consider a setting where the mediator does not have information advantage and the buyer's valuation is interdependent with the seller's private information. The mechanism may involve multi-round negotiations and flexible fee structures. We show that the mediator can restrict attention to a class of joint menu-selection mechanisms, where each mechanism can be represented as a two-dimensional menu. Each party privately selects an option from their own dimension and the two options together determine the menu entry. The mediator then recommends both parties whether to trade based on the jointly selected menu entry. We then establish an impossibility trilemma: no mechanism can simultaneously satisfy incentive compatibility, obedience, and informativeness. Motivated by this result, we characterize the optimal mechanisms under two relaxation conditions. First, when the seller's cost is constant, the optimal mechanism exhibits a threshold structure: trade occurs whenever the quality of the item exceeds a threshold that is decreasing in the buyer's type. Consequently, low-typed buyers receive more information, which in turn gives the mediator more power to charge from them. Second, when the mediator has veto power, the optimal mechanism also takes a threshold form, but in the opposite direction: trade occurs only if the quality falls below a threshold that is increasing in the buyer's type. As a result, items with lower qualities are more likely to be traded and the corresponding sellers benefit more, which discourages sellers of high qualities from participating and gives rise to a ``lemons market'' effect.


}%




\KEYWORDS{mediation mechanism, bilateral trade, interdependent valuations} 

\maketitle


\section{Introduction}\label{sec:Intro}
Mediators have become increasingly prevalent across modern marketplaces and play a crucial role in facilitating transactions. On the buyer side, searching for a suitable seller independently can be both time-consuming and costly, whereas working with a mediator provides immediate access to a curated pool of sellers. On the seller side, reaching potential buyers often requires marketing effort, and sometimes high advertising expenses. By relying on a mediator, sellers can rapidly connect with a targeted customer base and improve the likelihood of a successful match. A prominent example is the real estate market. According to statistics from National Association of Realtors for 2023–2024, 89\% of home buyers and 90\% of sellers in the United States completed their transactions with the assistance of a real estate agent or broker~\citep{theclose2024}. Moreover, 90\% of buyers reported that their broker was a helpful source of information, and 87\% of sellers said that they would definitely or probably recommend their agent for future services~\citep{nar2024}. Mediators are also indispensable in many other markets. In used-car markets, the report by Mordor Intelligence shows that over 55\% of used-car transactions in the U.S. involve third-party platforms such as CarMax and Carvana, and is projected to expand at a 2.86\% CAGR through 2030~\citep{MordorUSUsedCar2024}. In the market for mergers and acquisitions (M\&A), firms commonly rely on investment banks and advisory services for screening, valuation, and negotiation. According to \cite{BainGlobalMA2024}, global M\&A volume reached 3.5 trillion USD in 2024 and transaction volume is up 7\%. The continued growth in global M\&A activity underscores the critical role of financial Mediators. 

A key feature of these markets is the interdependence of buyers' valuations and the presence of information asymmetry. The buyers' valuation of the item is influenced by its quality, yet the quality is typically private information held by the seller. This asymmetry creates friction, making it difficult for buyers to accurately assess the item. Therefore, a trustworthy mediator becomes essential. For instance, in a used-car market, buyers cannot directly observe the true condition of a vehicle, including its maintenance history or hidden defects. Similarly, in mergers and acquisitions, a firm typically lacks full information about the target's financial health, operational capacity, or underlying risks before the M\&A took place. 

Another important feature is the dynamic interaction between the players and their autonomy in deciding whether to trade. Either party can choose not to proceed at any point before the transaction is finalized. Although the mediator has considerable flexibility in designing the trading mechanism, such as exchanging information or charging fees, the ultimate trading decision is made by the buyer and the seller, and the mediator cannot force a transaction to occur. This is clearly illustrated in the M\&A market: even though investment banks provide due diligence, valuation, and negotiation support, the acquiring and target firms independently evaluate the information and ultimately decide whether to complete the deal. 

In this paper, we study how the mediator should design a dynamic trading mechanism to maximize revenue when the buyer's valuation is interdependent. In particular, we consider a trade market with two players: a seller and a buyer. The seller owns an item and privately observes its quality, while the buyer has a private type, reflecting his preferences. The seller's valuation depends solely on the item's quality, whereas the buyer's valuation depends on both his type and the quality. The transaction is facilitated by an uninformed mediator. The mediator ex-ante designs a mechanism that specifies how the interaction unfolds. Within the mechanism, the mediator can privately communicate with each player to elicit or disclose information and may charge fees for providing mediation services. The interaction can proceed over multiple rounds, and at the end the mediator proposes transaction prices and a trade recommendation. Importantly, both players independently decide whether to trade and may opt out of the mechanism at any stage.

Given the mediator's extensive flexibility in mechanism design, directly characterizing a revenue-maximizing mechanism appears daunting. Our first result is that, for revenue maximization, it is without loss of generality to restrict attention to a class of \emph{joint menu-selection mechanisms}. Such a mechanism consists of a two-dimensional menu and two payment functions, one for each player. Each player privately selects an option from their corresponding dimension according to their own type. The two selected options jointly determine a menu entry (more precisely, a distribution), based on which the mediator issues a recommendation on whether to trade and charges each player according to their payment functions. This result implies that the mediator need not design complicated mechanisms involving multi-round communication, belief manipulation, or intricate fee schedules. Instead, it suffices to specify the two-dimensional menu and the prices associated with the options for each dimension.

Based on the simplified space, we formulate the mechanism design problem as an optimization program. Unlike the standard auction design model, our setting faces two novel challenges. First, the trade decision is made by the players themselves, which means we need to ensure that after receiving a recommendation, players get better payoffs from following the recommendation, which we call ``obedience''. It also makes the incentive compatibility (IC) constraint more involving (players may perform a so-called ``double deviation'' by misreporting and not following the recommendation). Second, in contrast to the mechanism design literature that provides incentives only to the buyer~\citep{alaei2024optimal,ashlagi2023sequential,zhou2025optimal}, in our setting, the mediator must simultaneously incentivize both the buyer and the seller.


We establish a fundamental impossibility trilemma: no mechanism can simultaneously achieve incentive compatibility, obedience, and informativeness. In other words, any incentive compatible and obedient mechanism must be uninformative: the mediator either always recommends trade or never recommends trade. We prove this result in three steps. First, we show that satisfying the seller's IC and obedience constraints requires the seller's payment to be constant. Second, restricting attention to informative mechanisms, we prove that given a constant seller payment, the buyer's IC constraints imply that the buyer's payment must also be constant. Finally, we demonstrate that a constant buyer payment cannot simultaneously satisfy the buyer's IC and obedience constraints. Motivated by this result, we introduce two relaxation conditions, under which we characterize the optimal mechanisms, respectively.


First, we show that when the seller's cost is constant, the optimal mechanism admits a threshold structure: the mediator recommends trade whenever the quality exceeds a carefully chosen threshold. This quality threshold is decreasing in the buyer's type, reflecting that buyers with higher types are more likely to trade. The optimal mechanism features several interesting properties. For example, the mediator fully extracts the seller's surplus, while the buyer has positive surplus that is strictly increasing in the buyer's type. Moreover, the buyer's payment function is decreasing in their type, implying that low-typed buyers are charged with higher prices, which contrasts with most results in existing literature. The intuition is that the low-typed buyers face a higher quality threshold, making the mediator's recommendation more informative for them, which in turn grants the mediator more power to extract these buyers' surplus. 

Second, we assume the mediator has veto power over the transaction. In other words, the players cannot complete trade without the consent of the mediator. Under this assumption, we show that the optimal mechanism again takes a threshold form, but in the opposite direction: the mediator recommends trade when the quality is below a type-dependent threshold, and the threshold is increasing in the buyer's type, indicating that high-typed buyers are less likely to trade. Moreover, sellers with higher qualities face lower trading probabilities and earn less surplus, which may discourage their participation and give rise to a ``lemons market'' effect.


From a technical perspective, our contributions are twofold. First, we derive the optimal mechanism in closed form despite the presence of nonlinear IC constraints arising from potential double deviations. In particular, we first relax the IC constraints by temporarily excluding double-deviation terms, which yields a tractable problem that can be solved using a Myersonian approach. In light of the decreasing-payment property, we further show that if double deviation is allowed, the buyer's most profitable deviation is to mimic the highest type and always trade. Leveraging this insight, we demonstrate that the solution to this relaxed problem remains feasible (and therefore optimal) in the original problem. Second, when considering general distributions that fail to satisfy the regularity condition, the classical ironing method of \cite{myerson1981optimal} cannot be applied directly. The difficulty arises from the interdependent nature of valuations in our model, which requires ironing an interdependent virtual value function. To address this challenge, we develop a modified ironing procedure tailored to our setting and use it to characterize the optimal mechanism in the general case.


\subsection{Related Literature}

Our paper is related to the literature on bilateral trade with a third party. \cite{myerson1983efficient} characterize the optimal mechanism that maximizes the broker's expected profit. Their game is static, where neither the buyer nor the seller can exit the mechanism once they choose to participate. In our setting, players can freely exit at any stage and make the trade decision, which renders the revelation principle based on Bayesian Nash equilibrium inapplicable. \cite{balseiro2022dynamic} study a two-sided market in which a platform intermediates repeated trade between a seller offering independent items and multiple buyers. They propose an approximate mechanism with provable profit guarantees. In contrast, we focus on single item transactions but allow for multi-round communications before the trade, and we fully characterize the optimal mechanism for the mediator. There is a much larger body of work that studies brokers who aim to maximize efficiency~\citep{vcopivc2016optimal, cesa2024bilateral, loertscher2023bilateral}. The seminal work by \cite{myerson1983efficient} characterize the optimal (second-best) mechanism, that is, the mechanism that maximizes gains from trade (GFT) among all implementable outcomes. Their mechanism denies trade when the buyer's valuation exceeds the seller's. \cite{eilat2021bilateral} assume the mediator cannot deny efficient trade and investigate how the mediator communicates with each party and the bounds on what they can achieve. \cite{zhang2019efficient} assume the broker is imperfectly informed about the buyer's and the seller's values and provide a condition under which efficient trade can be achieved. \cite{hajiaghayi2025gains} study whether the presence of a self-interested broker undermines the possibility of achieving a constant-factor approximation to the first-best GFT. They show that when the broker adopts a profit-maximizing mechanism, it guarantees a 1/2-approximation to the first-best GFT when the buyer's and seller's values follows uniform distributions. In contrast, the mediator in our model aims to maximize his own expected revenue. In addition, all above studies typically assume independent values between the buyer and the seller, while we consider an interdependent valuation setting, which is more general and expands the model's applicability.


Our paper is also related to the literature on bilateral trade with interdependent values. \cite{gresik1991ex} identifies sufficient conditions under which a deterministic solution exists and then proves an impossibility result, which extends the impossibility result of \cite{myerson1983efficient} to general valuations market. \cite{fieseler2003partnerships} derive conditions under which efficient trade is possible. \cite{dobzinski2025bilateral} consider a more general environment and ask when approximation is possible. These works study environments with two-sided incomplete information, in which both the buyer's and seller's valuations are interdependent. In contrast, \cite{jehiel2006partnership} consider a one-sided asymmetric information setting where the seller is fully informed while the buyer is uninformed of its value. They identify conditions under which full efficiency cannot be reached. \cite{kunimoto2022efficient} overcome this impossibility issue by employing a two-stage mechanisms. In our model, the seller is fully informed but the buyer is partial informed about his valuation. Besides, rather than focusing on efficiency, our mechanism is optimal in terms of profit maximizing.


Our paper builds on the information design literature pioneered by \cite{rayo2010optimal} and \cite{kamenica2011bayesian}, and is particularly related to recent work on information disclosure in mechanism design, where a seller can disclose additional information to the buyers~\citep{esHo2007optimal, bergemann2007information, li2017discriminatory, kolotilin2017persuasion, wei2024reverse, schottmuller2023optimal}. 
In contrast to these papers, the designer in our setting is a mediator who has no information about both players' private type. Thus, the mediator must incentivize both the buyer and the seller to reveal their private information. Moreover, the mediator cannot force the parties to trade: a transaction occurs only if both players consent. This introduces additional obedience constraints and creates the possibility of double deviations in the mechanism design.


Our paper also relates to the literature on information selling~\citep{babaioff2012optimal, bergemann2018design, chen2020selling}. \cite{liu2021optimal} consider an information seller who can design a sequential mechanism to sell information to an information buyer. They show that it is without loss of generality to focus on a one-round mechanism, and they characterize the optimal mechanism for selling the seller's information. In their information-selling setting, the buyer makes his decision after the information is sold, so the payment for information does not affect his decision since it is sunk. In contrast, in our setting, the mediator makes a profit only when both players choose to trade and the mediator earns the spread. This leads to substantial technical differences in determining obedience constraints, and consequently leads to different mechanism properties. Besides, we study a three-player environment, which necessitates using the solution concept of perfect Bayesian equilibrium (PBE).

The problem studied in this paper falls within the literature on communication equilibrium \citep{forges1993five, Bergemann2016InformationDB}, in which an uninformed mediator recommends actions to players based on the information elicited from them. In general settings, the number of constraints grows exponentially with the number of players, making the problem intractable. By focusing on a bilateral trade environment with only two players, each endowed with two possible actions, we are able to overcome this complexity and fully characterize the optimal mechanism.


The rest of the paper is organized as follows. Section \ref{sec:model} defines our model and the mechanism space, and show that the mediator can focus on the set of joint menu-selection mechanism without loss of generality. Section \ref{sec:main problem} formulate the mechanism design problem as an optimization program. In Section \ref{sec:impossibility}, we establish an impossibility trilemma. In Section \ref{sec:constant cost}, we characterize the optimal mechanism when the seller's cost is constant, under Monotone Hazard Rate condition. In Section \ref{sec:partial obedience}, we derive the optimal mechanism when the mediator has veto power, under regularity condition. Section \ref{sec:general} discuss how our results can be extended to more general distributions. Section \ref{sec:Conclusion} conclude, and the Appendix presents the technical proofs.

\section{Model}\label{sec:model}

\subsection{Setup}
Consider a trade market where a seller $s$ owns an item and wishes to sell it to a buyer $b$ through a mediator $me$. Let $q$ denote the quality of the item, which is privately observed by the seller, and let $t\in T$ represent the buyer's private type, capturing their preferences. We assume both $q$ and $t$ are random variables independently drawn from publicly known distributions $G(q)$ and $F(t)$, with $Q = [\underline{q}, \bar{q}]$ and $T = [\underline{t}, \bar{t}]$ being their supports. Suppose that both $G(q)$ and $F(t)$ are differentiable, with corresponding probability density functions $g(q)$ and $f(t)$.

Let $v(q, t)$ be the valuation of a buyer with type $t$ for an item of quality $q$. With slight abuse of notation, we use $v(t)$ to denote the buyer's expected valuation for the item under his prior beliefs about $q$:
\begin{equation*}
    v(t) = \int_{\underline{q}}^{\bar{q}} v(q, t) g(q) \, \dd q.
\end{equation*}
In practice, it is reasonable to assume that buyers prefer higher-quality items and that those with stronger preferences are willing to pay more for the same item. Accordingly, we assume that $v(q, t)$ is monotone increasing in both quality $q$ and type $t$. For expositional clarity, we will assume that the valuation is linear in $t$, i.e., there exists a real-valued function $\alpha(q) > 0, \alpha'(q) > 0$ such that $v(q, t) = \alpha(q) t$. The production cost of a seller with quality $q$ is denoted by $c(q)$. In practice, higher-quality items are typically more costly to produce. Therefore, we assume that the cost function is proportional to the quality, i.e., $c(q)=kq$ with $k> 0$. 

The buyer and the seller must rely on a mediator to facilitate the transaction. The mediator, who is initially uninformed about both players’ private information, ex-ante designs a mechanism that specifies how the transaction will be conducted. Within this mechanism, the mediator can communicate privately with each player to collect or disclose information and to impose fees. The interaction may involve multiple rounds, and both players are free to exit the mechanism at any stage. In the end, the mediator proposes a transaction price for each player. Each player then independently decides whether to trade, and the transaction occurs only if both players choose to do so. Our goal is to design a mechanism that maximizes the mediator's expected revenue.

\subsection{Mechanism Space and the Revelation Principle}
To design a revenue-maximizing mechanism, it is essential to first specify the mechanism space that the mediator can employ. 
Prior work by \cite{babaioff2012optimal} and \cite{liu2021optimal} study an information-selling problem and define a mechanism space that describes the direct interaction between a buyer and a seller. We extend this framework by introducing a mediator into the game. Specifically, we consider the following set of sequential mechanisms. 

\begin{definition}[Sequential Mechanism]
\label{def:general_mechanism}
A sequential mechanism is a mechanism that induces a finite extensive-form game between the seller and the buyer. The game consists of a finite number of $n$ communication rounds followed by one trade round. Each communication round is \emph{private}, meaning that the interaction between the mediator and one player is unobservable to the other. Let $H^k_{me} \in \mathcal{H}^k_{me}$ denote the interact history of the mediator at the end of $k$-th communication round. 
In the $k$-th communication round, the following three stages take place sequentially:
\begin{enumerate}
    \item Each player $i \in \{s, b \}$ sends a message $m_i^k \in M_i^k$ to the mediator. We allow $M_i^k$ to be a singleton, in which case player $i$ has no communication at this stage.
    \item The mediator sends a message $m^k_{me, i} \in M_{me, i}^k$ privately to each player $i \in \{s, b\}$. The mediator's messaging strategy is given by a function $S^k_{me,i}: \mathcal{H}^{k-1}_{me} \times M_b^k \times M_s^k \mapsto \Delta(M^k_{me, i}) $. Similarly, if $M^k_{me, i}$ is a singleton, this stage of communication is inactive.
    \item Each player $i \in \{s, b\}$ pays a fee $p_i^k \in \mathbb{R}$ to the mediator.
\end{enumerate}
In the final trade round:
\begin{enumerate}
    \item The mediator posts a price $P_i$ to each player $i \in \{s, b\}$ and announces trade or no trade. 
    \item The buyer and the seller independently decide whether to trade. The trade occurs if and only if both players agree to do so. 
    \item If the trade occurs, the buyer pays $P_b$ to the mediator, and the seller receives $P_s$ from the mediator. Otherwise, no trade occurs, and the mechanism terminates.
\end{enumerate}

At any stage, if any player $i\in {s, b}$'s action is not permissible by the mechanism, the mechanism terminates. Thus, players are free to exit at any time.
\end{definition}

Once the mediator announces the mechanism, the induced game is well-defined for both the seller and the buyer. We make the standard assumption that the mediator has commitment power and will behave exactly as committed.

\textbf{Preference.} We assume a quasi-linear preference for both players. Let $z$ be the terminal node that contains the entire history of players' actions. Denoted by $\tau_b(z)$ the total payment of the buyer along the history $z$, $\tau_s(z)$ the total payment the seller receives along the history $z$. Then the buyer's payoff and the seller's payoff can be written as: 
\begin{equation*}
\begin{aligned}
    u_b(z)
    &= v(q, t)\cdot \mathbf{1}_{\text{Trade}} - \tau_b(z), \\
    u_s(z)
    &= -c(q)\cdot \mathbf{1}_{\text{Trade}} + \tau_s(z).
\end{aligned}
\end{equation*}

\textbf{Solution Concept.} Our solution concept is perfect Bayesian equilibrium (PBE). To formally define it, we need to define the extensive-form game (between the buyer and the seller) induced by the mechanism. We relegate the definition of the game to Appendix A. Here we briefly define PBE. 

Let $H_i$ denote the set of non-terminal nodes where player $i\in \{s, b \}$ moves. $\mathcal{I}_i$ is a partition of $H_i$ such that an element $I \in \mathcal{I}_i$ is called an \emph{information set} of player $i$. Let $A(I)$ be the action set of the information set $I$. A \emph{strategy} $S_i(I)$ of player $i\in \{s, b \}$ is a function that assigns a probability distribution over $A(I)$ to each information set $I\in \mathcal{I}_i$. A \emph{belief} is a function $\mu_i: H_i \mapsto [0, 1]$ that assigns to each node $h$ a probability such that the probabilities of the nodes in any information set sum up to 1, i.e., $\sum_{h\in I} \mu_i(h)=  1, \forall I\in \mathcal{I}_i$, for both players $i\in \{s, b \}$.

An \emph{assessment} is a pair $(S, \mu)$, where $S$ is a strategy profile and $\mu$ is a belief profile. Let $Z_I$ be the set of all terminal nodes reachable from some nodes in $I$, and $\Pr(z \mid S,\mu, I)$ the conditional probability of reaching terminal node $z\in Z_I$ given an assessment $(S, \mu)$ and information set $I\in \mathcal{I}_i$. Then the conditional expected payoff of player $i$ given $(S, \mu)$ at information set $I$ is defined as $U_{i,I}(S \mid \mu) = \sum_{z\in Z_I} \Pr(z \mid S, \mu, I)u_i(z)$.

\begin{definition}[Perfect Bayesian Equilibrium]
An assessment $(S^*, \mu^*)$ is a weak perfect Bayesian equilibrium if the following conditions hold for each player $i\in \{s, b\}$.
\begin{itemize}
    \item \emph{Sequential rationality.} For every information set $I \in \mathcal{I}_i$ and every $S_i$, 
    \begin{equation*}
        U_{i,I}(S_i^*, S^*_{-i} \mid \mu^*) \ge U_{i,I}(S_i,S^*_{-i} \mid \mu^*).
    \end{equation*}
    \item \emph{Belief consistency.} For every information set $I \in \mathcal{I}_i$ and every $h \in I$,
    \begin{equation*}
        \mu^*(h) = \frac{\Pr(h\mid S^*)}{\sum_{h' \in I} \Pr(h'\mid S^*)},
    \end{equation*}
    where $\Pr(h' \mid S^*)$ is the probability of reaching node $h'\in I$ given $S^*$.
\end{itemize}
\end{definition}

\textbf{Revelation Principle.}
Deriving the revenue-maximizing mechanism is challenging due to the richness of the mechanism space. When designing optimal mechanisms for selling physical goods, the celebrated revelation principle~\citep{myerson1979incentive, gibbard1973manipulation} allows us to, without loss of generality, focus on direct and truthful mechanisms. However, in our setting, the mediator acts as an information hub, coordinating information exchange between the buyer and the seller, and there is no bound on the number of communication rounds that may occur.

Nevertheless, we show that our setting admits a stronger form of the revelation principle: it is without loss of generality to restrict attention to a class of \emph{joint menu-selection mechanisms}. A joint menu-selection mechanism is composed of both a two-dimensional menu and two payment functions. Each player privately selects an option from their own dimension and the mediator recommend both parties whether to trade or not based on the menu entry that is jointly determined by the selected options.

\begin{definition}[Joint Menu-Selection Mechanism]
A joint menu-selection mechanism consists of a signal set $M = \{ \text{trade, no trade} \}$, a signaling rule $\pi: Q \times T \mapsto \Delta(M)$, a payment function for the buyer $P_b: T \mapsto \mathbb{R}$ and a payment function for the seller $P_s: Q\mapsto \mathbb{R}$. The mechanism proceeds as follows:
\begin{enumerate}
    \item Each player privately reports their type $q'\in Q$ and $t'\in T$ to the mediator.
    \item The mediator sends a signal according to $\pi(q', t')$.
    \item Each player decides whether to trade at the suggested prices $P_b(t')$ and $P_s(q')$.
    \item If both players choose to trade, the transaction occurs, the buyer pays $P_b(t')$ and the seller receives $P_s(q')$. Otherwise, no trade occurs and the mechanism terminates.
\end{enumerate}
\end{definition}

\begin{remark}
The entire process of the mechanism can equivalently be represented as a menu selection process. The buyer is offered a set of choices $\{ P_b(t), \pi(\cdot, t)  \}_{t\in T}$, where each choice specifies a price $P_b(t)$ and an associated dimensional option $\pi(\cdot,t)$. The seller is offered an analogous set $\{P_s(q), \pi(q,\cdot)\}_{q\in Q}$. After each player privately makes a choice decision, the two options jointly determine a menu entry $\pi(q, t)$ (i.e., a distribution), based on which the mediator sends a signal.

\end{remark}

Without loss of generality, we use $\pi(q, t)$ to denote the probability of sending signal ``trade''. Naturally, the probability of sending signal ``no trade'' is $1-\pi(q, t)$.


In a joint menu-selection mechanism, each signal can be viewed as an action recommendation. We say that a joint menu-selection mechanism is \emph{feasible} if both players are willing to participate (IR), report their types truthfully (IC), and follow the mediator's recommendation (Obedience). The following result shows that it is without loss of generality to focus on joint menu-selection mechanisms.

\begin{proposition}[Revelation Principle]
\label{prop:revelation}
For any sequential mechanism and a corresponding weak PBE $(S^*, \mu^*)$, a joint menu-selection mechanism with 
\begin{equation*}
\begin{aligned}
    \pi(q, t)
    & \coloneqq \Pr(\text{trade occurs} \mid S^*, q, t), \\
    P_b(t)
    & \coloneqq 
    \frac{\E_{q\sim g} \!\left[ \E_z \!\left[p_b(z)\mid S^*, q, t\right] \right]}
         {\E_{q\sim g} \!\left[\pi(q, t)\right]}, \\
    P_s(q)
    & \coloneqq
    \frac{\E_{t\sim f} \!\left[ \E_z \!\left[p_s(z)\mid S^*, q, t\right] \right]}
         {\E_{t\sim f} \!\left[\pi(q, t)\right]} .
\end{aligned}
\end{equation*}

is feasible and achieves the same expected revenue for the mediator and the same expected payoff for both players.
\end{proposition}

In any game induced by a sequential mechanism, a perfect Bayesian equilibrium must satisfy both IR and IC, otherwise, the prescribed strategy profile would not form an equilibrium. Moreover, once a trade decision is made, neither player can benefit from taking the opposite action. Therefore, when constructing the direct recommendation mechanism, we simply replicate these trade outcomes by having the mediator send ``trade'' signal, which must satisfy the obedience requirement.

This result is reminiscent of the classical revelation principle, but differs in a key respect. The standard revelation principle in~\cite{myerson1982optimal} applies to static mechanisms under Bayesian Nash equilibrium, whereas our setting is dynamic. Because the mediator's multi-round communication with one side is unobservable to the other, the appropriate solution concept here is perfect Bayesian equilibrium rather than BNE.

\section{Optimization Formulation of the Mechanism Design Problem}\label{sec:main problem}
Based on the simplified mechanism space, our goal is to design a joint menu-selection mechanism that maximizes the mediator's revenue:
\begin{equation*}
    \int_{\underline{q}}^{\bar{q}} \int_{\underline{t}}^{\bar{t}} \pi(q, t) [P_b(t) - P_s(q)] f(t) g(q) \,\dd t \dd q,
\end{equation*}
subject to the following constraints:
\begin{itemize}
    \item \textbf{Individual Rationality.} For each player, truthful participation in the mechanism (truthful reporting and following the recommendation) yields an expected payoff at least that from opting out.
    \item \textbf{Obedience.} Conditional on truthful reporting, following the mediator's recommendation is optimal for each player.
    \item \textbf{Incentive Compatibility.} Truthful reporting is optimal for each player. \endnote{In some of mechanism design literature, incentive compatibility requires that truthful reporting and following recommendation jointly constitute an optimal strategy. In our setting, obedience is imposed as a separate constraint, which guarantees that following the recommendation is optimal after truthful reporting. Therefore, incentive compatibility can be defined solely in terms of truthful reporting without loss of generality.}
\end{itemize}


Now we analyze these constraints one by one to formulate the above problem as mathematical program.

\textbf{IR constraints.} 
Both players' payoff is 0 if they do not participate. Therefore, the expected payoff of a buyer of type $t$, reporting truthfully and following the recommendation, is
\begin{equation}
\label{eq:buyer utility}
\begin{aligned}
    U_b(t)
    &= \mathbb{E}_{q} [ \pi(q,t)(v(q,t) -P_b(t)) ] \\
    &= \int_{\underline{q}}^{\bar{q}} \pi(q, t) [ v(q, t) - P_b(t) ] g(q)\, \dd q .
\end{aligned}
\end{equation}

To ensure the buyer's participation in the mechanism, the following IR constraint is required
\begin{align}
\label{eq:buyer IR}
    U_b(t) \ge 0, \forall t \in T. \tag{IR-Buyer}
\end{align}
Similarly, the expected payoff of a seller with quality $q$, reporting truthfully and following the recommendation, is
\begin{equation}
\label{eq:seller utility}
\begin{aligned}
    U_s(q) 
    &= \E_{t} [\pi(q,t) ( P_s(q) - c(q)) ]  \\ 
    &=\int_{\underline{t}}^{\bar{t}} \pi(q, t) [ P_s(q) - c(q) ]  f(t)  \,\dd t .
\end{aligned}
\end{equation}

The IR constraint on the seller's side is given by
\begin{align}
\label{eq:seller IR}
    U_s(q)  \ge 0, \forall q\in Q.  \tag{IR-Seller}
\end{align}

\textbf{Obedience constraints.} 
To ensure the obedience of the buyer, we need to pose the following two constraints: 
\begin{equation}
\label{buyer ob 1}
    \int_{\underline{q}}^{\bar{q}} g(q \mid t,\text{trade}) v(q, t)\, \dd q - P_b(t) \ge 0, \forall t \in T .
\end{equation}

\begin{equation}
\label{buyer ob 0}
    \int_{\underline{q}}^{\bar{q}} g(q \mid t, \text{no trade}) v(q, t)\, \dd q- P_b(t) \le 0, \forall t \in T .
\end{equation}
The first constraint ensures that when receiving ``trade'' signal, the buyer's expected payoff from trade is at least 0. Similarly, the second constraint ensures the obedience of ``no trade'' signal. Once receiving signal ``trade'', a buyer of type $t$ updates their belief over $q$ via Bayes' rule:
\begin{equation}
\label{eq:belief update}
    g(q \mid t, \text{trade}) 
    = \frac{\pi(q, t) g(q) }{ \int_{\underline{q}}^{\bar{q}} \pi(q', t) g(q') \, \dd q' }.
\end{equation}
Based on Equation \eqref{eq:belief update}, Constraint \eqref{buyer ob 1} is equivalent to
\begin{equation*}
    \int_{\underline{q}}^{\bar{q}} \pi(q, t) [ v(q, t) - P_b(t)  ] g(q)  \,\dd q \ge 0, \forall t \in T.
\end{equation*}
Using similar arguments, we can rewrite Constraint \eqref{buyer ob 0} as
\begin{equation*}
	\int_{\underline{q}}^{\bar{q}} \pi(q, t)[v(q, t) - P_b(t)] g(q)\,\dd q\ge  v(t)- P_b(t), \forall t\in T.
\end{equation*}
Therefore, we can summarize the obedience constraint for the buyer as follows:
\begin{align}
\label{Ob-Buyer}
    \int_{\underline{q}}^{\bar{q}} \pi(q, t)[v(q, t) - P_b(t)] g(q)\,\dd q \ge  \max \{0, v(t) - P_b(t)  \}, \forall t\in T. \tag{OB-Buyer}
\end{align}

For the seller, the obedience constraint for a ``trade'' recommendation is given by
\begin{equation*}
    \int_{\underline{t}}^{\bar{t}} \pi(q, t) [P_s(q) - c(q)] f(t) \, \dd t \ge 0.
\end{equation*}
And the obedience constraint for a ``no trade'' recommendation is
\begin{equation*}
    \int_{\underline{t}}^{\bar{t}} [1-\pi(q, t)] [P_s(q) - c(q)] f(t)\, \dd t \le 0.
\end{equation*}
Combining the two, the seller's obedience constraint can be written compactly as
\begin{equation}
\label{ob seller}
    \int_{\underline{t}}^{\bar{t}} \pi(q, t) [P_s(q) - c(q)] f(t) \, \dd t  \ge  \max\{ 0, P_s(q) - c(q) \}, \forall q\in Q.
\end{equation}
\begin{remark}
Each player's IR constraint is implied by their corresponding obedience constraints. To some extent, obedience can be interpreted as a kind of ex-post IR, guaranteeing that players obtain non-negative payoffs conditional on every realized signal.
\end{remark}

Constraint \eqref{ob seller} can be further simplified by case analysis:
\begin{enumerate}
    \item If $P_s(q) - c(q) < 0$, then the right-hand side of Constraint \eqref{ob seller} is 0. This inequality holds if and only if $\pi(q, t) = 0 , \forall t \in T$, meaning the mediator always makes ``no trade'' recommendation.
    \item If $P_s(q) - c(q) > 0$, then the right-hand side of Constraint \eqref{ob seller} becomes $P_s(q) - c(q)$. The inequality holds if and only if $\pi(q, t) = 1, \forall t\in T$, i.e., the seller is always recommended to trade.
    \item If $P_s(q) - c(q) = 0$, the obedience constraint always holds. The seller is indifferent, so any recommendation satisfies obedience.
\end{enumerate}
Define  $Q_+ \coloneqq  \{ q\mid P_s(q) > c(q) \}$ be the seller-profitable region, $Q_{-} \coloneqq  \{ q \mid P_s(q) < c(q) \}$ be the seller-unprofitable region and $Q_{=} \coloneqq \{ q\mid P_s(q) = c(q) \}$ be the seller-indifferent region. The above analysis implies that the obedience constraint imposes a tight relationship between $P_s$ and $\pi$:
\begin{align}
\label{Ob-seller}
    \pi(q, t) \in \begin{cases}
        \{1\} \quad \text{for all } t\in T & \text{if } q\in Q_+,\\
        \{0\} \quad \text{for all } t\in T & \text{if } q\in Q_{-},\\
        [0,1] \quad \text{for all } t\in T & \text{if } q\in Q_{=}.\\
    \end{cases} \tag{OB-Seller}
\end{align}
In other words, obedience requires that the mediator refrain from discriminating across buyers in seller-profitable region and seller-unprofitable region. Only in seller-indifferent region may the mediator discriminate against buyers in terms of information.

\textbf{IC constraints.} 
The derivation of the IC constraints is more involved because a player who misreports may face an induced experiment $\pi$ that is no longer obedient for them. We begin by analyzing the buyer's IC constraint. When receiving signal ``trade'', the expected payoff of a buyer with type $t$ who misreports $t'$ and follows the recommendation is given by
\begin{equation*}
	U_b(t'; t) = \int_{\underline{q}}^{\bar{q}} \pi(q, t') [v(q, t) - P_b(t')] g(q)\,\dd q.
\end{equation*}
The buyer can also choose not to trade if that benefits them since $\pi(q, t')$ may not be obedient for them. Therefore, the actual expected payoff is $\max \{U_b(t'; t) , 0  \}$. Similarly, when receiving signal ``no trade'', the buyer's expected payoff from disobeying the recommendation is 
\begin{equation*}
    \int_{\underline{q}}^{\bar{q}} [1- \pi(q, t')] [v(q, t) - P_b(t')]  g(q) \,\dd q
     = v(t) - P_b(t') - U_b(t'; t).
\end{equation*}
And the actual payoff is $\max \{0, v(t) - P_b(t') - U_b(t'; t)  \}$. Combining both cases, the expected payoff of a buyer with type $t$ from misreporting type $t'$ is $\max \{ U_b(t'; t), 0 \} + \max \{ 0, v(t) - P_b(t') - U_b(t'; t) \}$. Therefore, the IC constraint for the buyer becomes
\begin{equation}
\label{IC buyer}
	U_b(t) \ge \max \{ U_b(t'; t), 0 \}  + \max \{ 0, v(t) - P_b(t') - U_b(t'; t) \}, \forall t, t' \in T.
\end{equation}

Although this constraint appears complicated at first glance, we can simplify it by leveraging our previous results:
\begin{enumerate}
	\item If $t> t'$, we have $U_b(t'; t) > U_b(t'; t') \ge 0$, where the first inequality follows from the assumption $\alpha(q) >0$ and the second comes from the obedience of type $t'$. In this case, the right-hand side of Constraint \eqref{IC buyer} becomes $\max \{ U_b(t'; t), v(t) - P_b(t') \}$.
	\item If $t < t'$, we have $v(t) - P_b(t') - U_b(t'; t) < v(t') -P_b(t') - U_b(t';t') \le 0 $ for similar reasons. In this case, the right-hand side of Constraint \eqref{IC buyer} becomes $\max \{U_b(t';t), 0 \}$. Note that $U_b(t)\ge 0$ is already implied by the IR constraint. Therefore, the remaining constraint is $U_b(t) \ge U_b(t'; t)$.
\end{enumerate}
To summarize, the IC constraint for the buyer can be reduced to the following:
\begin{align}
\label{Simple IC buyer}
	U_b(t) \ge \max_{t'\in T} \{ U_b(t'; t), v(t) - P_b(t') \}, \forall t \in T,  \tag{IC-Buyer}
\end{align}
where $v(t) - P_b(t')$ denotes the buyer's payoff from double deviation. 

Similarly, for the seller, when receiving ``trade'' signal, the expected payoff of a seller with quality $q$ who misreports $q'$ and chooses to trade is:
\begin{equation*}
	U_s(q'; q) =  \int_{\underline{t}}^{\bar{t}} \pi(q', t) [P_s(q') - c(q)] f(t) \,\dd t.
\end{equation*}
The true payoff in this case is $\max \{ U_s(q';q), 0 \}$. And when receiving ``no trade'' signal, the true expected payoff is $\max \{0, P_s(q') - c(q) - U_s(q';q) \}$.
Combining both cases, the expected payoff of a seller with quality $q$ from misreporting $q'$ is $\max \{ U_s(q'; q), 0 \} + \max \{ 0, P_s(q') - c(q) - U_s(q';q) \}$. Therefore, the IC constraint for the seller becomes
\begin{equation}
\label{IC seller}
    U_s(q) \ge \max \{ U_s(q'; q), 0 \} + \max \{ 0, P_s(q') - c(q) - U_s(q';q) \}, \forall q, q'\in Q.
\end{equation} 
Similarly, we can also simplify Constraint \eqref{IC seller} based on previous results:
\begin{enumerate}
    \item If $q' \in Q_{-}$, we have $U_s(q'; q) = 0$ because of the obedience constraint. In this case, the right-hand side of Constraint \eqref{IC seller} becomes $\max\{0, P_s(q') - c(q) \} $. Note that $U_s(q) \ge 0$ is already implied by the IR constraint. Therefore, the only non-redundant constraint is $U_s(q) \ge P_s(q') - c(q) $.
    \item If $q' \in Q_+$, we have $U_s(q'; q) = P_s(q') - c(q)$ and $P_s(q') - c(q) - U_s(q'; q) = 0$ for the same reason. In this case, the right-hand side of Constraint \eqref{IC seller} is also $\max\{0, P_s(q') - c(q)\}$. And the only non-redundant constraint is also $U_s(q) \ge P_s(q') - c(q) $.
    \item If $q' \in Q_{=}$, we continue to discuss two different conditions:
    \begin{enumerate}
        \item If $q' > q$, we have $U_s(q'; q) > 0$ and $P_s(q') - c(q) - U_s(q';q) >0$. In this case, the IC Constraint \eqref{IC seller} is also $U_s(q) \ge P_s(q') - c(q)$.
        \item If $q' < q$, we have $U_s(q'; q) <0$ and $P_s(q') - c(q) - U_s(q';q) <0$. In this case, the IC Constraint \eqref{IC seller} becomes $U_s(q) \ge 0$, which is already implied by the IR constraint.
    \end{enumerate}
\end{enumerate}
To summarize, the IC constraint for the seller can be finally reduced to the following:
\begin{align}
\label{Simple IC seller}
    U_s(q) \ge \max_{q'\in Q} P_s(q') - c(q), \forall q\in Q, \tag{IC-Seller}
\end{align}
where $P_s(q') - c(q)$ denotes the seller's payoff from double deviation.

Given the above analysis, the optimal mechanism design problem can be formulated as the following optimization program, with functional variables $\pi(q, t), P_b(t)$ and $P_s(q)$:
\begin{maxi}
{}   
{\int_{\underline{q}}^{\bar{q}} \int_{\underline{t}}^{\bar{t}} \pi(q, t)[P_b(t) - P_s(q)]f(t)g(q) \,\mathrm{d}t \mathrm{d}q}  
{\label{eq:LP}}  
{}  
\addConstraint{ \, \eqref{Ob-Buyer},  \eqref{Simple IC buyer}  }{ }{}
\addConstraint{ \eqref{Ob-seller}, \eqref{Simple IC seller} }{ }{}
\end{maxi}

\section{The Impossibility Trilemma}\label{sec:impossibility}
In this section, we obtain an impossibility trilemma that no mechanism can simultaneously satisfy IC, obedience and informativeness. In other words, an IC and obedient mechanism must be uninformative, i.e., it must always (with probability 1) recommend ``trade'' or never (with probability 0) recommend ``no trade''. We begin by deriving several structural properties that any feasible mechanism must satisfy. The following lemmas characterize the key restrictions imposed on the seller's payment function.

\begin{lemma}
\label{lem:impossible 1}
If a mechanism satisfies the seller's IC constraints, then the seller-indifferent region $Q_{=}$ must contain at most one element.
\end{lemma}

If the mediator pays the seller exactly the cost at two different qualities, then a low-quality seller would deviate by misreporting the high quality and receive a payment that exceeds their true cost.
From a geometric perspective, the IC constraints require that function $P_s(q)$ and $c(q)$ intersect at most once. 

\begin{lemma}
\label{lem:impossible 2}
If a mechanism satisfies the seller's obedience and IC constraints, then the seller's payment function $P_s(q)$ must satisfy:
\begin{enumerate}
    \item The payment in the seller-profitable region should be constant and no less than that in other regions, i.e., for any $q_1, q_2 \in Q_+$ and any $q_3 \notin Q_{+}$, 
    \begin{equation*}
        P_s(q_1) = P_s(q_2)  \ge P_s(q_3).
    \end{equation*}
    \item The payment should be upper bounded by the minimum cost in the region outside the seller-profitable region. More precisely, for any $q'\in Q$, $q\notin Q_{+}$,
    \begin{equation*}
        P_s(q') \le c(q).
    \end{equation*}
\end{enumerate}
\end{lemma}
By the obedience constraint, a seller with a quality in $Q_+$ always choose to trade and their payoff is $P_s(q) - c(q)$. The first part follows from the fact that if payments differed within $Q_+$, a seller would simply misreport a quality that yields a higher payment. And for the second part, if $P_s(q') >  c(q)$, a seller with $q$ will misreport to $q'$ to obtain a positive payoff $P_s(q') - c(q)$. 

Let $P_+$ denote the constant payment on $Q_+$, i.e., $P_s(q) = P_+$ for all $q\in Q_+$. If $P_+ \ge c(\bar{q})$ or $P_+ \le c(\underline{q})$, the mediator either always recommends trade or never recommends trade. In both cases, the signal conveys no information to the players. We refer to such mechanisms as uninformative. Next, we focus on informative mechanisms where $c(\underline{q}) < P_+ < c(\bar{q})$. According to the definition of $c(q)$, there exists a unique $q_\ell \in (\underline{q}, \bar{q})$ such that $c(q_\ell) = P_+$. Based on Lemma \ref{lem:impossible 2}, we can characterize the sets $Q_+$ and $Q_{-}$ as follows.

\begin{corollary}
\label{cor:cutoff_Q_sets}
If a mechanism satisfies the seller's obedience and IC constraints, then the sets $Q_{+}$ and $Q_{-}$ admit the following cutoff structure:
\begin{enumerate}
\item If $Q_= \neq\emptyset$, then $P_s(q)$ and $c(q)$ intersect at point $(q_\ell, P_+) $, and
\begin{equation*}
    Q_+=[\underline q,q_\ell),\qquad Q_==\{q_\ell\},\qquad Q_-=(q_\ell,\bar q].
\end{equation*}
\item If $Q_==\emptyset$, then 
\begin{equation*}
    Q_+=[\underline q,q_\ell)\qquad\text{and}\qquad Q_-=[q_\ell,\bar q].
\end{equation*}
\end{enumerate}
\end{corollary}

From a revenue-maximizing perspective, the mediator can set $P_s(q_\ell) = P_+$ and set $\pi(q, t) = 0 $ if he find it is unprofitable. 
According to the obedience constraint and the structure of $P_s(q)$, once choosing $P_+$, the signaling rule $\pi$ is pinned down to $\pi^{q_\ell}$:
\begin{equation}
\label{eq:pi q_l}
    \pi^{q_\ell}(q, t) \in \begin{cases}
        \{1\} \quad \text{for all } t\in T & \text{if } q \in [\underline{q}, q_\ell),\\
        [0,1] \quad \text{for all } t\in T & \text{if } q = q_\ell,\\
        \{0\} \quad \text{for all } t\in T & \text{if } q \in  (q_\ell, \bar{q}].
    \end{cases}
\end{equation}

Next, we characterize the feasible region of buyer payment function $P_b(t)$, under $\pi^{q_\ell}$. We restate the feasibility constraints for the buyer as follows:
\begin{equation}
\label{eq:feasibility buyer}
    U_b(t) \ge \max_{t'\in T} \{U_b(t'; t), v(t) - P_b(t'), 0 \}, \forall t\in T.
\end{equation}

\begin{lemma}
\label{lem:IC property}
Constraint $U_b(t) \ge \max_{t'\in T} \{ U_b(t'; t), 0 \}, \forall t\in T$ can be fully characterized by
\begin{align}
	&R^{\pi}_b(t) \text{ is non-decreasing in $t$}  \label{eq:monotone} \\
	&U_b(t) = U_b(\underline{t} ) + \int_{\underline{t} }^t R_b^{\pi}(x) \,\dd x \label{eq:IC property} \\
	&U_b(\underline{t}) \ge 0  \label{eq:IR property}
\end{align}
where $R^{\pi}_b(t) = \int_{\underline{q}}^{\bar{q}} \alpha(q) \pi(q, t) g(q)\,\dd q$.
\end{lemma}
Lemma~\ref{lem:IC property} follows directly from the standard envelope theorem. Note that the monotonicity condition \eqref{eq:monotone} is analogous to the allocation monotonicity condition in classical auction design. The weighted term $\alpha(q)$ is caused by the interdependent value function $v(q,t)$. Under $\pi^{q_\ell}$, we have 
\begin{equation*}
    R^{\pi^{q_\ell}}_b(t) = \int_{\underline{q}}^{q_\ell} \alpha(q) g(q) \,\dd q,
\end{equation*}
which is constant in $t$, satisfying the monotonicity constraint \eqref{eq:monotone}. For convenience, we denote this quantity by $R^{q_\ell}$. Next, we characterize the conditions under which constraint \eqref{eq:IC property} is satisfied.

\begin{lemma}
\label{lem:buyer constant}
Under $\pi^{q_\ell}$, constraint \eqref{eq:IC property} is satisfied if and only if $P_b(t) = P_b(t'), \forall t, t'\in T$.
\end{lemma}

From a geometric perspective, $R^{\pi}_b(t)$ represents the slope of the buyer's utility function $U_b(t)$ at $t$. Under $\pi^{q_\ell}$, this slope is constant over $T$, implying that $U_b(t)$ must be linear in $t$. The linearity of $U_b(t)$, in turn, requires the buyer payment to be independent of type $t$.

Now we are ready to characterize the set $P_b(t)$ that satisfies the feasibility constraint \eqref{eq:feasibility buyer}.
\begin{lemma}
\label{lem:buyer feasibility}
Under $\pi^{q_\ell}$, the buyer feasibility constraint \eqref{eq:feasibility buyer} is satisfied if and only if the buyer is charged a constant payment $P_b$ and
\begin{equation*}
    \bar{t}\cdot \E\left[\alpha(q)\mid q\ge q_\ell\right]
    \;\le\; P_b \;\le\;
    \underline{t}\cdot \E\left[\alpha(q)\mid q\le q_\ell\right].
\end{equation*}
\end{lemma}

Intuitively, the payment cannot be too high, since otherwise, the buyer with the lowest type receives negative utility, violating individual rationality \eqref{eq:IR property}. At the same time, if $P$ falls below the lower bound, always trading yields strictly higher utility than following the mediator's recommendation for the buyer with the highest type, violating the obedience constraint. 

However, one can easily verify that, for any $q_\ell \in (\underline{q}, \bar{q} )$, the lower bound on the buyer payment is strictly larger than the upper bound. Consequently, the feasible set is empty for any such $q_\ell$. When $q_\ell \le \underline{q}$ or $q_\ell \ge \bar{q}$, the mechanism degenerates to an uninformative one: the mediator either always recommends trade or never recommends trade. Therefore, we have the following result.

\begin{theorem}[Impossibility Trilemma]
\label{the:impossible}
No mechanism can simultaneously satisfy the following three properties: IC, obedience and informativeness. And there exists two sets of uninformative mechanisms that satisfy the first two properties:
\begin{itemize}
    \item Always-trade mechanism: $\pi(q, t) = 1, \forall t\in T, q\in Q$, with $P_+ \ge c(\bar{q})$, $P \le v(\underline{t})$.
    \item Never-trade mechanism: $\pi(q, t) = 0, \forall t\in T, q\in Q$, with $P_+ \le c(\underline{q})$, $P\ge v(\bar{t})$.
\end{itemize}
\end{theorem}

Under the always-trade mechanism, the mediator earns positive revenue if and only if $v(\underline{t}) \ge c(\bar{q})$. In contrast, the never-trade mechanism yields zero revenue for the mediator.


\section{Optimal Mechanism under Constant Cost}\label{sec:constant cost}
The above impossibility trilemma indicates that any IC and obedient mechanism must be trivial. Therefore, in this section, we consider a relaxed condition and revisit the mechanism design problem when the seller's cost is constant.

\begin{assumption}[Constant Cost]
The seller's cost function is constant, i.e., $c(q) = c, \forall q\in Q$.
\end{assumption}

This assumption is consistent with markets in which quality is privately observed but the seller's cost is fixed. Such a constant-cost assumption is widely adopted in the mechanism design literature~\citep{wei2024reverse,alaei2024optimal,guo2025designing,zhou2025optimal}.


\begin{assumption}[Monotone Hazard Rate (MHR)]
\label{MHR}
Distribution $F(t)$ has monotone hazard rate, if 
$\frac{f(t)}{1- F(t)}$ is monotone non-decreasing in $t$.
\end{assumption}
MHR distributions are widely studied in the mechanism design literature \citep{cai2011extreme} and holds for many commonly used distributions, such as the uniform, normal and exponential distributions. We will discuss how to relax this assumption in Section \ref{sec:MHR}.


\subsection{Optimization Formulation under Constant Cost}

According to Lemma \ref{lem:impossible 2}, the seller's payment should be upper bounded by the constant cost. Then we have following result.
\begin{corollary}
For any feasible mechanism, $P_s(q)$ must satisfy: for any $q\in Q$, $P_s(q) \le c$.
\end{corollary}


From a revenue-maximizing perspective, we can set $P_s(q) = c$ for all $q\in Q$ without loss of generality. \endnote{If $P_s(q) = c$ is unprofitable in certain regions, the mediator can set $\pi(q, t) = 0$ for those non-profitable region.} The feasibility constraints on the buyer side remain unchanged. Therefore, the optimization problem reduces to:
\begin{maxi}
{}
{ \int_{q} \int_{t} \pi(q, t) [P_b(t) - c] f(t) g(q)\,\dd t\dd q }
{\label{eq:orignal LP}}
{}
\addConstraint{U_b(t)}{\ge 0 \quad}{\forall t \in T}
\addConstraint{ U_b(t) }{\ge\max_{t'\in T} \{U_b(t'; t), v(t) - P_b(t') \} \quad}{\forall t \in T}
\end{maxi}



\subsection{Characterization of the Optimal Mechanism}
Program \eqref{eq:orignal LP} involves IC constraints for double deviation, which renders that the standard Myersonian approach cannot be applied directly.
We solve Program \eqref{eq:orignal LP} in two steps. First, we relax the IC constraint to
\begin{equation}
\label{eq:relaxed IC}
    U_b(t) \ge U_b(t'; t), \forall t, t'\in T,
\end{equation}
and use Myersonian approach to solve the relaxed problem. Second, we prove that the optimal solution to the relaxed problem satisfies the original constraints, thus remains optimal for the original problem.

\subsubsection{The Relaxed Problem}
We write the relaxed problem as follows:
\begin{maxi}
{}
{\int_{\underline{q}}^{\bar{q}} \int_{\underline{t}}^{\bar{t}} \pi(q, t) [P_b(t) - c] f(t) g(q)\,\dd t\dd q }
{\label{eq:relaxed LP}}
{}
\addConstraint{U_b(t)}{\ge 0, \quad}{\forall t \in T}
\addConstraint{U_b(t)}{\ge U_b(t'; t), \quad}{\forall t, t'\in T}
\end{maxi}


Based on Lemma \ref{lem:IC property}, we reformulate the objective function.

\begin{lemma}
\label{lem:rewrite obj}
The objective of Program \eqref{eq:relaxed LP} can be rewritten as follows:
\begin{equation}
	- U_b(\underline{t}) + \int_{\underline{q}}^{\bar{q}} \int_{\underline{t}}^{\bar{t}} \pi(q, t)  \eta(q, t) f(t) g(q) \,\dd t\dd q, \label{eq:rewrite obj}
\end{equation}
where $\eta(q, t) = \alpha(q) \Big[ t - \frac{1- F(t)}{f(t)} \Big] - c $.
\end{lemma}

We note that the revenue depends only on $\pi$ and is not affected by the detailed payment function. Thus, the key step in determining the optimal mechanism is to characterize the structure of $\pi$, which is governed by function $\eta(q, t)$. We now analyze its properties.

\begin{lemma}
\label{lem:eta single root}
Given any $t$, there exists at most one $q_0\in Q$ such that $\eta(q_0, t) = 0$.
\end{lemma}

Based on the above property, we introduce the following threshold function:
\begin{equation*}
    \lambda(t) = \begin{cases}
        \underline{q} & \text{if } \eta(q, t) > 0, \forall q\in Q, \\
        \bar{q} & \text{if }  \eta(q, t) < 0,  \forall q\in Q, \\
        q_0 & \text{otherwise}. 
    \end{cases}
\end{equation*}
Intuitively, $\lambda(t)$ represents the cutoff quality level at which the mediator becomes indifferent between recommending trade and no trade for a buyer of type $t$. 

We are now ready to present the optimal solution to the relaxed problem.
\begin{theorem}
\label{prop:relax optimal}
The following mechanism is an optimal solution to Program \eqref{eq:relaxed LP}:
\begin{equation*}
    \pi^*(q, t) = \begin{cases}
	1 & \text{if } q \ge \lambda(t), \\
	0 & \text{otherwise} ,
    \end{cases}
\end{equation*}
\begin{equation}
\label{eq:optimal buyer pay}
	P_b^*(t) = \begin{cases}
		 \E [ v(q, t) |q \ge \lambda(t)] - \frac{\int_{t_1}^t R_b^{\pi^*}(x)\,\dd x}{1- G(\lambda(t))} & \text{if } t\ge t_1, \\
		v( \bar{q} , t_1)  & \text{otherwise},
	\end{cases}
\end{equation}
where $t_1 \coloneqq \inf_T \{t: 1 - G( \lambda (t)) > 0 \}  $ denotes the lowest type such that a buyer will be recommended to trade with positive probability.
\end{theorem}

In the optimal mechanism, a buyer purchases the item only when its quality exceeds a type-dependent threshold $\lambda(t)$, which is decreasing in the buyer's type. As a result, buyers with higher types trade with higher probability and obtain larger surplus. Although buyers with lower types receive more informative signals, their equilibrium surplus is smaller. In other words, an optimal mechanism features both price and information discrimination. On the seller side, the mechanism always pays the constant cost and fully extracts the surplus, leaving sellers with zero payoff from participation.

\begin{example}
If $G$ and $F$ follow uniform distribution on $[1, 2]$, $v(q, t) = q t$ where $\alpha(q) = q$, and $c(q)= 2$, then we have $\eta(q, t) = q(2t - 2) - 2$. In the optimal mechanism, the mediator recommends trade if $q \ge \lambda(t) = \frac{1}{t-1}$.

There are two cases for the buyer:
\begin{itemize}
    \item If $t < 1.5$, then $q< \frac{1}{t-1}, \forall q \in [1, 2]$. The mediator will never recommend trade, and the buyer's payment function is $P_b^*(t) = 3$.
    \item If $t\ge 1.5$, the mediator will recommend trade when $q\ge   \frac{1}{t-1} $. The buyer's payment function $P_b^*(t)$ can be calculated as follows
    \begin{equation*}
        P_b^*(t) = \frac{ \int_{\frac{1}{t-1}}^2 q t \,\dd q - \int_{1.5}^t \int_{\frac{1}{x-1}}^2 q \, \dd q \dd x }{2-\frac{1}{t-1}} = 2 + \frac{1}{2(t-1)}.
    \end{equation*}
\end{itemize}

\end{example}


\subsubsection{The Original Problem}
Now we return to the original problem and show that $(\pi^*, P_b^*)$ stays feasible for Program \eqref{eq:orignal LP}. We first provide a structural property of the payment function $P_b^*(t)$. Define $t_2 \coloneqq \sup_{T} \{ t: 1 - G(\lambda(t)) < 1 \}$, then buyers with type between $t_1$ and $t_2$ will receive a trade recommendation with probability strictly between 0 and 1. 

\begin{lemma}
\label{prop:reverse}
$P_b^*(t)$ is decreasing in $t$. In particular, $P_b^*(t)$ is strictly decreasing on $(t_1, t_2)$, and constant on $[t_2, \bar{t}]$.
\end{lemma}
In contrast with most existing studies, the buyer's payment function is decreasing in $t$: buyers with higher types pay less. The intuition comes from the decreasing threshold $\lambda(t)$. Upon receiving a trade recommendation, the buyer's posterior belief about $q$ is simply the prior belief restricted on the interval $[\lambda(t), \bar{q}]$. Hence, lower-typed buyers face higher thresholds, making the recommendation more informative, which in turn grants the mediator more power to charge from lower-typed buyers.


Next, we show that $(\pi^*, P_b^*)$ satisfies constraints $U_b(t) \ge v(t) - P_b^*(t')$ that were omitted in the relaxation.
By Lemma \ref{prop:reverse}, it is sufficient to check 
\begin{equation*}
    U_b(t) \ge v(t) - P_b^*(\bar{t}), \forall t\in T,
\end{equation*}
which corresponds to the deviation where a buyer of type $t$ misreports the highest type $\bar{t}$ and then always chooses to trade. When the highest type $\bar{t}$ is recommended to trade with probability 1, this deviation is equivalent to ``reporting $\bar{t}$ and always following the recommendation'', which is already ruled out as unprofitable by the relaxed IC condition \eqref{eq:relaxed IC}. The proof in the appendix extends this argument to the general case where $\bar{t}$ may not always receive a trade recommendation.

\begin{theorem}
\label{prop:origin optimal}
$(\pi^*, P_b^*)$ is an optimal solution to original problem, described in Program \eqref{eq:orignal LP}.
\end{theorem}
The MHR condition serves two distinct roles in the proof of Theorem \ref{prop:origin optimal}. First, it guarantees that the trading probability $R_b^{\pi}(t)$ is increasing in $t$, playing a role analogous to the standard regularity condition in the mechanism design literature. Second, it implies a decreasing pricing structure, which in turn ensures that no buyer has an incentive to perform a double deviation by misreporting their type and always choosing to trade.

\section{Optimal Mechanism with a Veto-power Mediator}\label{sec:partial obedience}
In this section, we derive the optimal mechanism when the mediator has veto power. The results in this section is based on the following conditions.

\begin{assumption}[Veto-power Mediator]
The mediator has veto power over the transaction: trade cannot occurs if the mediator disapproves.

\end{assumption}

This assumption is consistent with many real-world markets. For example, in mergers and acquisitions, an investment bank can withhold necessary approvals or documentation, effectively preventing the deal from closing.

\begin{assumption}[Regularity]
Define $\phi_b^-(t) = t - \frac{1- F(t)}{f(t)}$ and $\phi_s^+(q)  =q + \frac{G(q)}{g(q)}$.
We say a problem is regular if $\phi^-_b(t)$ is non-decreasing in $t$ and $\frac{\phi^+_s(q)}{\alpha(q)}$ is non-decreasing in $q$.
\end{assumption}

The monotonicity of $\phi_b^-(t)$ is the standard regularity condition in the mechanism design literature~\citep{myerson1981optimal}. This condition is strictly weaker than the MHR condition and is satisfied by a substantially broader class of distributions. The monotonicity of $\phi_s^+(q)$ is also commonly imposed in literature~\citep{myerson1983efficient, jehiel2006partnership}. By contrast, the second condition, which requires monotonicity of the quality-adjusted term $\frac{\phi_s^+(q)}{\alpha(q)}$, is specific to our interdependent-value environment. We will discuss the irregular cases in Section \ref{sec:irregular}.

\subsection{Optimization Formulation with a Veto-power Mediator}
In this setting, when the mediator sends a ``no trade'' signal, obedience is automatically satisfied since trade cannot occur. Following the analysis in Section \ref{sec:main problem}, we obtain the following simplifications. First, the IR constraints remain unchanged. Second, the obedience constraints are equivalent to the IR constraints. Third, the buyer's IC constraints reduce to
\begin{equation}
\label{eq:partial IC buyer}
    U_b(t) \ge U_b(t';t),  \forall t, t' \in T.
\end{equation}
And the IC constraints for the seller reduces to
\begin{equation}
\label{eq:partial IC seller}
    U_s(q) \ge U_s(q';q), \forall q, q' \in Q.
\end{equation}

Therefore, we can formulate the mechanism design problem as the following program:
\begin{maxi}
{}   
{\int_{q} \int_{t} \pi(q, t)[P_b(t) - P_s(q)]  \,\dd F(t) \dd G(q)}  
{\label{partial ob LP}}  
{}  
\addConstraint{U_b(t)}{\ge \max_{t'\in T} \{ U_b(t';t), 0 \} \quad}{\forall t \in T}
\addConstraint{U_s(q)}{\ge \max_{q'\in Q} \{ U_s(q';q), 0\} \quad}{\forall q \in Q}
\end{maxi}

Compared with traditional one-sided mechanism design problems, the central challenge in our setting is that the mediator must design incentives for both the buyer and the seller simultaneously.

\subsection{Characterization of the Optimal Mechanism}

We first characterize the feasible solutions of Program \eqref{partial ob LP}. 
\begin{lemma}
\label{lem:partial IC property}
A mechanism $(\pi, P_b, P_s)$ satisfies all the constraints in Program \eqref{partial ob LP} if and only if:
\begin{align}
    & R^{\pi}_b(t) \text{ is non-decreasing  in $t$}, \label{eq:partial buyer_monotone}\\
    & R^{\pi}_s(q) \text{ is non-increasing in $q$}, \label{eq:partial seller_monotone}\\
    & U_b(t) = U_b(\underline{t}) + \int_{\underline{t}}^t R^{\pi}_b(x) \,\dd x, \label{eq:partial IC property buyer} \\
    & U_s(q) = U_s(\underline{q}) - k\int_{\underline{q} }^q R^{\pi}_s(x) \,\dd x, \label{eq:partial IC property seller}\\
    & U_b(\underline{t} ) \ge 0, \label{eq:partial IR property buyer}\\
    & U_s(\bar{q}) \ge 0, \label{eq:partial IR property seller}
\end{align}
where $R_s^{\pi} (q) = \int_{\underline{t}}^{\bar{t}}  \pi(q, t) f(t) \,\dd t$.
\end{lemma}
The characterization of buyer-side feasibility constraints follows directly from Lemma \ref{lem:IC property}. Interestingly, we find that in any feasible mechanism, the seller's trading probability must be decreasing in $q$.  Constraint \eqref{eq:partial IC property seller} further implies that the seller's surplus is a decreasing and convex function of $q$. Consequently, to guarantee the IR of the seller, it suffices to ensure that the seller with the highest quality obtains a non-negative payoff.

Based on Lemma \ref{lem:partial IC property}, we can rewrite the objective function in Program \eqref{partial ob LP}.
\begin{lemma}
\label{lem:partial equal revenue}
The objective function of Program \eqref{partial ob LP} can be rewritten as:
\begin{equation}
\label{eq:partial equal revenue}
    -U_b(\underline{t} )-U_s(\bar{q} ) +\int_{\underline{q}}^{\bar{q}} \int_{\underline{t}}^{\bar{t}}\pi(q, t)[\alpha(q) \phi^-_b(t) - k \phi^+_s(q)] f(t)g(q) \,\dd t \dd q.
 \end{equation}
\end{lemma}
Similar to Lemma \ref{lem:rewrite obj}, the mediator's revenue depends only on the signaling rule $\pi$ and is not affected by the specific payment functions. As a result, the mediator's revenue can be maximized pointwisely by setting $\pi(q, t) = 1$ for $q$ and $t$ pairs with $\alpha(q) \phi_b^-(t)  \ge k \phi_s^+(q) $.

\begin{remark}
The term $\alpha(q) \phi_b^-(t)$ admits a natural interpretation as an \emph{interdependent virtual value}. It scales the buyer's independent virtual value $\phi_b^-(t)$ by the quality-dependent factor $\alpha(q)$ in the buyer's value function. Under this interpretation, the signaling rule recommends trade whenever the interdependent virtual value weakly exceeds the seller's virtual cost $k\phi_s^+(q)$.
\end{remark}


We now present the optimal mechanism that solves Program \eqref{partial ob LP}.

\begin{theorem}
\label{thm:optimal_mechanism}
The following mechanism is an optimal solution to Program \eqref{partial ob LP}:
\begin{equation*}
    \pi^*(q, t) = \begin{cases}
	1 & \text{if} \quad \phi_b^-(t) \ge \frac{k\phi_s^+(q)}{\alpha(q)}, \\
	0 & \text{otherwise} . 
    \end{cases}
\end{equation*}
\begin{align}
    P_b^*(t) = &\mathbb{E}_q [v(q, t) \mid \pi^*(q,t) = 1] - \frac{\int_{\underline{t}}^{t} R_b^{\pi^*} (x) \,\dd x }{ \mathbb{E}_q[\pi^*(q, t)]} ,\label{eq:opt_buyer_payment}\\
    P_s^*(q) =& \mathbb{E}_t[c(q) \mid \pi^*(q, t) = 1] + \frac{\int_{q}^{\bar{q}} R_s^{\pi^*}(x)\, \dd x }{\mathbb{E}_t[\pi^*(q, t)]} . \label{eq:opt_seller_payment}
\end{align}
\end{theorem}

In the above optimal mechanism, a buyer trades only for items whose quality falls below a type-dependent threshold, and this threshold is increasing in the buyer's type. Consequently, buyers with higher types obtain larger surplus. Relatively, items with lower qualities are traded with higher probability, leading to the fact that sellers with lower qualities benefit more. This discourages sellers with high qualities from participating in the market and gives rise to a ``lemons market'' outcome \citep{akerlof1970market}.





\begin{example}
\label{example}
If $G$ and $F$ follow uniform distribution on $[1, 2]$, $v(q,t) = qt$ where $\alpha(q) = q$, and $c(q) = 1.5 q$ with $k=1.5$, then we have
$\phi^-_b(t) = t - \frac{1-F(t)}{f(t)} = 2t-2, \phi^+_s(q) = q+\frac{G(q)}{g(q)} = 2q-1$. 
In the optimal mechanism, the mediator will recommend them to trade if $2t - 2\ge \frac{ 3q-1.5 }{q}$, or equivalently, $t\ge 2.5 - \frac{1.5}{2q}$. There are two cases for the buyer:
\begin{itemize}
    \item If $t < 1.75$, then $t < 2.5-\frac{1.5}{2q},\forall q$, which means that the mediator will never recommend trade.
    \item If $t \ge 1.75$, the mediator will recommend trade when $q \le \frac{1.5}{5-2t}$. The buyer's payment function can be computed as follows
    \begin{equation*}
        P_b^*(t) = \frac{\int_1^{\frac{1.5}{5 - 2t} }qt \,\dd q  -\int_{1.75}^t \int_1^{\frac{1.5}{5 - 2x} } q \, \dd q\dd x}{\frac{1.5}{5 - 2t} - 1}  = \frac{1}{2}+\frac{15}{ 8(5-2t)}.
    \end{equation*} 
\end{itemize}

For the seller, there are also two cases.
\begin{itemize}
    \item If $q > 1.5$, then $2.5 - \frac{1.5}{2q} > t, \forall t$, which means the mediator will never recommend trade. 
    \item If $q \le 1.5$, the mediator will recommend trade when $t \ge 2.5 - \frac{1.5}{2q}$. The payment function for the seller can be computed as follows
    \begin{equation*}
        P_s^*(q) = \frac{\int_{2.5 - \frac{1.5}{2q}}^2 1.5 q \,\dd t +  \int_{q}^{1.5} \int_{2.5 - \frac{1.5}{2x}}^2 1.5 \,\dd t \dd x}{ \frac{1.5}{2q} -0.5 } = \frac{9q}{6-4q} \ln\left(\frac{3}{2q}\right).
    \end{equation*} 
    For $q=1.5$, we can take the limit and get $P_s^*(q)=2.25$.
\end{itemize}
\end{example}




\section{Generalizations}\label{sec:general}
Our results in Section \ref{sec:constant cost} and Section \ref{sec:partial obedience} rely on the monotone hazard rate condition and regularity condition, respectively. In this section, we discuss how these assumptions can be relaxed.

\subsection{Relaxing the MHR Assumption}\label{sec:MHR}
As discussed earlier, the MHR assumption is sufficient to ensures that $\eta(q, t)$ is increasing in $t$ and to rule out profitable double deviations. However, our approach does not depend on the MHR condition. Next, we relax this assumption (but still assuming monotone virtual value), and provide alternative conditions under which our approach still applies.

Recall that constraint $U_b(t) \ge v(t) - P_b(t'), \forall t, t'\in T$ precludes buyers from engaging in double deviations. We first establish that the function $U_b(t) - v(t)$ is monotone decreasing in $t$.

\begin{proposition}
\label{prop:decreasing}
The function $U_b(t) - v(t)$ is monotone decreasing in $t$.
\end{proposition}
Therefore, it suffices to ensure that the buyer with the highest type $\bar{t}$ has no incentive to perform a double deviation. When $\eta(\underline{q},\bar{t}) \ge 0$, our results still hold if and only if $\min_{t\in T} P^*_b(t) = P^*_b(\bar{t})$. This condition is necessary because buyer with type $\bar{t}$ trades with probability is 1. Hence, to prevent double deviations, type $\bar{t}$ must face the lowest payment. The condition is also sufficient, because any buyer who considers double deviation will optimally report the highest type $\bar{t}$. Since type $\bar{t}$ is recommended to trade with probability 1, ``reporting $\bar{t}$ and always trading'' is equivalent to ``reporting $\bar{t}$ and always following the recommendation'', and we have already ensured that the latter is unprofitable.

\subsection{Relaxing the Regularity Assumption}\label{sec:irregular}



For ease of presentation, we define $\psi(t) = \phi_b^-(t)$ and $\varphi(q) = \frac{k\phi_s^+(q)}{\alpha(q)}$. The revenue equation \eqref{eq:partial equal revenue} can be rewritten as $-U_b(\underline{t}) - U_s(\bar{q}) + \int_{\underline{q}}^{\bar{q}} \int_{\underline{t}}^{\bar{t}}\pi(q, t)\alpha(q)[\psi(t) - \varphi(q) ]\,\dd F(t)\dd G(q)$.

When the regularity condition fails to hold, the classical approach is to \emph{iron} the non-monotone virtual value function \citep{myerson1981optimal}. A formal definition of the ironing technique is provided in the appendix for completeness. We argue that the standard ironing procedure can only be applied to the function $\psi(t)$, but \textbf{not} to $\varphi(q)$ because the latter contains the additional scaling term $\alpha(q)$. To overcome this difficulty, we propose a modified ironing technique tailored to our setting, and use it to derive the optimal mechanism.


\begin{theorem}
\label{thm:optimal_mechanism_irregular}
For any irregular case, a mechanism with the following $\pi^*$ and the same payment function in Theorem \ref{thm:optimal_mechanism}  is feasible and maximizes revenue.
\begin{equation*}
    \pi^*(q, t) = \begin{cases}
	1 & \text{if } \bar{\psi}(t) \ge \bar{\varphi}(q), \\
	0 & \text{otherwise}.
    \end{cases}
\end{equation*}
where $\bar{\psi}(t)$ and $\bar{\varphi}(q)$ are the ironed functions.
\end{theorem}

The key difference between our ironing method and the standard Myerson ironing lies in the choice of measure on the quality space. In the classical setting, Myerson ironing proceeds by defining $w = G(q)\in [0, 1]$ and performing ironing on $\varphi(G^{-1}(w))$ in the $w$-space. In our setting, however, function $\varphi(q)$ is weighted by $\alpha(q)$, and the relevant measure becomes $\alpha(q) \dd G(q)$. A simple observation is that this weighted measure can be written as
\begin{equation*}
    \alpha(q) \dd G(q) = \dd w(q),
\end{equation*}
where $w(q) = \int_{\underline{q}}^q \alpha(x) g(x) \,\dd x$. This motivates  ironing in the $w(q)$-space rather than in the standard $G(q)$-space. 




\section{Conclusion}\label{sec:Conclusion}
This paper studies how an uninformed mediator should design a trading mechanism to maximize revenue in a bilateral trade with interdependent valuations. We first show that, despite the mediator's extensive flexibility in information exchange and fee design, it is without loss of generality to focus on joint menu-selection mechanisms. Within this class, we establish a fundamental impossibility trilemma: no mechanism can simultaneously satisfy incentive compatibility, obedience, and informativeness. Any mechanism satisfying the former two must be uninformative, offering no meaningful guidance to the parties. Motivated by this impossibility result, we introduce two relaxation conditions. When the seller's cost is constant, the revenue-maximizing mechanism takes a threshold form in which buyers with higher types face lower thresholds and buyers with lower types receive more informative recommendations. The mediator fully extracts the seller's surplus and extracts buyer surplus in a type-dependent manner. When the mediator has veto power, the optimal mechanism again exhibits a threshold structure but in the opposite direction, recommending trade only when quality is sufficiently low. Besides, sellers with lower qualities benefit more from the mechanism, which gives rise to a ``lemons market'' effect.

Our findings highlight both the power and the limitations of mediators in markets with interdependent values. These results clarify the precise conditions under which a mediator can profitably influence trade decisions and illustrate how revenue-maximizing designs redistribute surplus among market participants. These insights enrich the broader understanding of information, incentives, and strategic interaction in mediated markets.

\begingroup \parindent 0pt \parskip 0.0ex \def\enotesize{\normalsize} \theendnotes \endgroup




\bibliographystyle{informs2014} 
\bibliography{sample} 


\clearpage
%
%
%

\begin{APPENDICES}

\section{Omitted Definitions}
\subsection{Extensive-Form Game Induced by a Sequential Mechanism}

A sequential mechanism induces a finite extensive-form game. As the mediator has commitment power, the induced game is played between the seller and the buyer, and we can view the mediator's actions as chance moves. Formally, given a sequential mechanism, a corresponding two-player extensive-form game can be defined as a tuple $\Gamma = \langle  N, T, H, Z, \mathcal{H}, A, \mathcal{I}, u \rangle$, where

\begin{itemize}
    \item $N = \{b, s\}$ is the set of players;
    \item $T$ is a game tree, obtained by adding a \emph{nature} node to the tree described by $M$, where nature chooses the types of players at the beginning of the game;
    \item $H$ is a set of non-terminal nodes;
    \item $Z$ is a set of terminal nodes;
    \item $\mathcal{H} = \{H_0, H_s, H_b\}$ is a partition of $H$, where $H_0$ is the set of chance move nodes, and $H_s$ and $H_b$ are the sets of nodes where the seller and the buyer moves, respectively;
    \item $A$ is a function that maps a non-terminal node $h\in H$ to a set of available actions $A(h)$;
    \item  $\mathcal{I}= \{\mathcal{I}_i\}_{i\in N}$ is a collection of information partitions, where each $\mathcal{I}_i$ is a partition of $H_i$. An element $I \in \mathcal{I}_i$ is called an \emph{information set} of player $i$. Every information set $I$ satisfies $A(h) = A(h'), \forall h, h' \in I$. Thus for any information set $I\in \mathcal{I}_i$, we denote by $A(I)$ the set $A(h)$ for all $h \in I$;
    \item $u = (u_i)_{i\in N}$ is a collection of payoff functions, where each $u_i:Z \mapsto \mathbb{R}$ is the payoff function of player $i$.
\end{itemize}

In such a game tree, any node in $T$ can be uniquely determined by the path from the root to the node itself. Such a path is also called the history of the node. Thus from now on, we abuse notation and use $h$ or $z$ to denote both a node and its history.

\subsection{Ironing}

\begin{definition}[Ironing]
    Let $\psi(t)$ be any non-monotone function. 
\begin{enumerate}
	\item Define a random variable $w = F(t)$ and let
	\begin{equation*}
		h_b(w) = \psi(F^{-1}(w)),
	\end{equation*}
	where $F^{-1}(w)$ is the inverse function of $F(t)$. Note that $F(t)$ is continuous and strictly increasing since we assume that the density function $f(t)$ is always strictly positive. Thus the inverse function $F^{-1}$ is also continuous and increasing.
	\item Let $H_b:[0, 1] \mapsto \mathbb{R}$ be the integral of $h_b(w)$:
	\begin{equation*}
		H_b(w) = \int_0^w h_b(x) \,\dd x.
	\end{equation*}  
	\item Let $L_b:[0, 1]\mapsto \mathbb{R}$ be the convex envelope of the function $H_b$:
	\begin{equation*}
		L_b(w) = \min \{ \lambda H_b(w_1) + (1-\lambda)H_b(w_2) \},
	\end{equation*}
	where $\lambda, w_1, w_2 \in [0, 1]$ and $\lambda w_1 + (1-\lambda)w_2 =w$.
	\item Define $l_b:[0, 1] \mapsto \mathbb{R}$ such that
	\begin{equation*}
		l_b(w) = L_b'(w).
	\end{equation*}
	\item Then we obtain the ironed function $\bar{\psi}$:
	\begin{equation*}
		\bar{\psi}(t) = l_b(w) = l_b(F(t)).
	\end{equation*}
\end{enumerate}
\end{definition}

\section{Proof of Technical Results}

\subsection{Proof of Proposition \ref{prop:revelation}}

\begin{proof}{Proof.}
It suffices to prove that the constructed direct recommendation mechanism gives the buyer the same expected utility, as the case for the seller and mediator is similar.

We first derive the buyer's expected payoff in the direct recommendation mechanism, if both players truthfully report their types and follow the recommendation. Let $g(q \mid t, m = \text{trade})$ be the buyer's posterior belief over $q$ upon receiving signal ``trade''. We have
\begin{equation*}
    g(q \mid t, m = \text{trade}) = \frac{\pi(q,t)g(q)}{ \int_q\pi(q,t)g(q)\,\dd q}.
\end{equation*}
Therefore, when receiving signal ``trade'', the buyer's expected utility from trade is
\begin{align*}
    &\E_{q\sim g(q\mid t, m = \text{trade})}\left[v(q, t)\right]-P_b(t)\\
    =&\frac{\int_q\pi(q,t)g(q)v(q,t)\,\dd q-\E_{q\sim g} [ \E_z [p_b(z) \mid S^*, q, t] ] }{\int_q\pi(q,t)g(q)\,\dd q}.
\end{align*}
When receiving signal ``no trade'', the buyer's utility is 0. Therefore, the expected utility of the buyer with type $t$ is
\begin{align*}
    &\int_q \pi(q, t) g(q) \,\dd q \cdot \Big( \E_{q\sim g(q\mid t, m = \text{trade})}\left[v(q,t)\right]-P_b(t) \Big)\\
    =&\int_q\pi(q,t)g(q)v(q, t)\,\dd q-\E_{q\sim g} [ \E_z [p_b(z) \mid S^*, q, t] ]\\
    =& \E_{q\sim g} [ \pi(q, t) v(q, t) - \E_z [p_b(z) \mid S^*, q, t] ] \\
    =&\E_{q\sim g} [ Pr(\text{trade occurs} \mid S^*, q, t) v(q, t) - \E_z [p_b(z) \mid S^*, q, t] ] ,
\end{align*}

where the last term is the buyer's expected utility in the original sequential mechanism.

Now we show that this direct recommendation mechanism is feasible. Players shall truthfully report and follow the recommendation in the direct recommendation mechanism since $(S^*, \mu^*)$ forms a PBE in the original sequential mechanism, i.e., any player has no incentive to mimic another type or disobey the recommendation in the sequential mechanism. The IR constraint holds since players are willing to participate in the original sequential mechanism.
\hfill \Halmos
\end{proof}


\subsection{Proof of Lemma \ref{lem:impossible 1}}

\begin{proof}{Proof.}
We prove it by contradiction. Assume that there exist two distinct qualities $q_1 < q_2$ such that 
\begin{equation*}
    P_s(q_1) = c(q_1), \quad P_s(q_2) = c(q_2).
\end{equation*}
Therefore, a seller with these two qualities obtains 0 from participating the mechanism, i.e., $U_s(q_1) = U_s(q_2) = 0$. According to constraints \eqref{Simple IC seller}, the IC constraints for a seller with quality $q_1$ requires 
\begin{equation*}
    U_s(q_1) \ge  P_s(q') - c(q_1), \forall q'\in Q.
\end{equation*}
Plugging $U_s(q_1) = 0$ and choosing $q' = q_2$ yields
\begin{equation*}
    0 \ge P_s(q_2) - c(q_1) = c(q_2) - c(q_1).
\end{equation*}
This inequality does not hold because $c(q)$ is strictly increasing in $q$ and $q_2 > q_1$. Therefore, a seller with quality $q_1$ has incentive to misreport $q_2$. The contradiction shows that the set $\{ q\mid P_s(q) = c(q) \}$ must contain at most one element.
\hfill \Halmos
\end{proof}

\subsection{Proof of Lemma \ref{lem:impossible 2}}

\begin{proof}{Proof.}
Recall the seller's IC constraint
\begin{equation*}
    U_s(q) \ge P_s(q') - c(q), \forall q, q'\in Q,
\end{equation*}
where $U_s(q)$ denotes the seller's expected payoff from participating the mechanism when his true type is $q$. We discuss three cases $q\in Q_+$, $q\in Q_{-}$ and $q\in Q_{=}$ separately.
\begin{enumerate}
    \item When $q\in Q_+$: by the obedience constraint we must have $\pi(q, t) = 1$ for all $t\in T$. Hence, the seller's expected payoff when truthful is
    \begin{align*}
        U_s(q) &= \int_{t} \pi(q, t) [P_s(q) - c(q)]  f(t)\,\dd t \\
        &= P_s(q) - c(q).
    \end{align*}
    Plugging this expression into the IC constraint gives, for every $q'\in Q$,
    \begin{equation*}
        P_s(q) - c(q) \ge P_s(q') - c(q), 
    \end{equation*}
    which simplifies to
    \begin{equation*}
        P_s(q) \ge P_s(q').
    \end{equation*}
    Since the inequality holds for any pair $q, q'\in Q_+$, we must have $P_s(q_1) = P_s(q_2)$ for all $q_1, q_2 \in Q_+$. Combined with the previous display, this common value on $Q_+$ is no less than $P_s(q')$ for any $q' \notin Q_+$. This proves statement (1).
    \item When $q\in Q_{-}$: obedience implies $\pi(q, t)=0$ for all $t\in T$, hence $U_s(q) = 0$. Plugging it into the IC constraint yields, for every $q'\in Q$,
    \begin{equation*}
        0 \ge P_s(q') - c(q),
    \end{equation*}
    or equivalently
    \begin{equation*}
        P_s(q') \le c(q), \forall q'\in Q.
    \end{equation*}
    \item When $q\in Q_{=}$: we have $P_s(q) = r(q)$, hence $U_s(q) = 0$. Plugging it into the IC constraint yields
    \begin{equation*}
        P_s(q') \le c(q), \forall q'\in Q.
    \end{equation*}
    Combining the last two cases, we have 
    \begin{equation*}
        P_s(q') \le c(q), \forall q \notin Q_{+}, q'\in Q,
    \end{equation*}
    which proves statement (2).
\end{enumerate}
This completes the proof.
\hfill \Halmos
\end{proof}

\subsection{Proof of Corollary \ref{cor:cutoff_Q_sets}}

\begin{proof}{Proof.}
We first show that $Q_+\subseteq [\underline q,q_\ell)$. Indeed, take any $q\in Q_+$. By definition, $P_s(q)>c(q)$, and by Lemma~\ref{lem:impossible 2}, we have $P_s(q)=P_+$. Hence $P_+>c(q)$, which implies $q<q_\ell$. Therefore $Q_+\subseteq [\underline q,q_\ell)$.

Next we show that $[\underline q,q_\ell)\subseteq Q_+$. Fix any $q<q_\ell$, so $c(q)<c(q_\ell)=P_+$. Suppose, toward a contradiction, that $q\notin Q_+$. Then either $q\in Q_=$ or $q\in Q_-$.  If $q\in Q_=$, then $P_s(q)=c(q)<P_+$, contradicting Lemma~\ref{lem:impossible 2}, which states that payment should be upper bounded by the cost on $Q_{=}$. Similarly, if $q\in Q_-$, then $P_s(q)<c(q)<P_+$. By Lemma~\ref{lem:impossible 2}, payments on $Q_+$ should be less than $c(q)$, otherwise a seller of type $q$ can profitably deviate by reporting any $q' \in Q_+$ and receiving $P_+$, yielding strictly positive payoff $P_+ - c(q) > 0$. Thus, neither case is possible, and we must have $q \in Q_+$. Hence, $[\underline q,q_\ell)\subseteq Q_+$, and therefore
\begin{equation*}
    Q_+=[\underline q,q_\ell).
\end{equation*}

We now characterize $Q_=$ and $Q_-$. Since $Q$ is partitioned into $Q_+$, $Q_=$, and $Q_-$ and we have already shown $Q_+=[\underline q,q_\ell)$, it remains to classify qualities in $[q_\ell,\bar q]$. For any $q>q_\ell$, we have $c(q)>c(q_\ell)=P_+$. By Lemma~\ref{lem:impossible 2}, $P_s(q)\le P_+$, and thus $P_s(q)-c(q)\le P_+-c(q)<0$, implying $q\in Q_-$. Hence $(q_\ell,\bar q]\subseteq Q_-$.

Finally, consider $q=q_\ell$. If $Q_=\neq\emptyset$, then by the fact that $Q_+=[\underline q,q_\ell)$ and $(q_\ell,\bar q]\subseteq Q_-$, the only remaining point that can belong to $Q_=$ is $q_\ell$. If $Q_= =\emptyset$, then $q_\ell$ cannot satisfy $P_s(q_\ell)=c(q_\ell)$, and since $P_s(q_\ell)\le P_+=c(q_\ell)$, it follows that $P_s(q_\ell)<c(q_\ell)$, i.e., $q_\ell\in Q_-$. Therefore $Q_-=[q_\ell,\bar q]$.

This proves both statements.
\hfill \Halmos
\end{proof}




\subsection{Proof of Lemma \ref{lem:IC property}}

\begin{proof}{Proof.}
We first show the necessity of Lemma \ref{lem:IC property}. We can rewrite $U_b(t'; t)$ as follows:
\begin{align*}
	U_b(t'; t) &= U_b(t') + (t- t') \int_{q\in Q} \alpha(q) \pi(q, t') g(q) \,\dd q  \\
	&= U_b(t') + (t-t') R_b^{\pi}(t').
\end{align*}
So the IC constraint becomes:
\begin{equation*}
	U_b(t) \ge U_b(t') + (t-t') R_b^{\pi}(t').
\end{equation*}
Switching $t$ and $t'$, we obtain:
\begin{equation*}
	U_b(t') \ge U_b(t) + (t' - t) R_b^{\pi}(t).
\end{equation*}
Combining these two inequalities, we have:
\begin{equation*}
	(t-t') R_b^{\pi}(t') \le U_b(t) - U_b(t') \le (t-t') R_b^{\pi}(t).
\end{equation*}
Note that $(t-t') (R_b^{\pi}(t) - R_b^{\pi}(t')) \ge 0$, which implies that $R_b^{\pi}(t)$ is non-decreasing in $t$.

If $t > t'$, we can divide the above inequality by $t -t'$ and get:
\begin{equation*}
	R_b^{\pi}(t') \le \frac{U_b(t) - U_b(t') }{t-t'} \le R_b^{\pi}(t').
\end{equation*}
When $t$ approaches $t'$, we have:
\begin{equation*}
	\frac{\dd U_b(t)}{\dd t} =R_b^{\pi}(t).
\end{equation*}
This analysis still holds if $t < t'$. Therefore, we get:
\begin{equation*}
	U_b(t) = U_b(\underline{t}) + \int_{\underline{t}}^t R_b^{\pi}(x) \,\dd x.
\end{equation*}

We observe that $U_b(t)$ is non-decreasing in $t$ since $R_b^{\pi}(t) \ge 0$ by definition. So function $U_b(t)$ is minimized at $\underline{t}$ and IR constraint is equivalent to $U_b(\underline{t}) \ge 0$.

Next, we show the sufficiency of Lemma \ref{lem:IC property}. We can rewrite the IC constraint as follows:
\begin{equation*}
	U_b(t) \ge U_b(t') + (t- t') R_b^{\pi}(t').
\end{equation*}
Constraint \eqref{eq:IC property} indicates IC constraints, because when $t> t'$, we have:
\begin{equation*}
	U_b(t) - U_b(t') = \int_{t'}^t R_b^{\pi}(x) \,\dd x \ge (t- t') R_b^{\pi}(t').
\end{equation*}
A similar analysis holds when $t < t'$. By definition $R_b^{\pi}(t)\ge 0, \forall t$, together with constraint \eqref{eq:IC property} and $U_b(\underline{t})\ge 0$ imply that $U_b(t) \ge 0$, for all $t$.
\hfill \Halmos
\end{proof}

\subsection{Proof of Lemma \ref{lem:buyer constant}}

\begin{proof}{Proof.}
Under $\pi^{q_\ell}$, constraint \eqref{eq:IC property} reduces to
\begin{equation*}
    U_b(t) = U_b(t') + (t- t') R^{q_\ell}.
\end{equation*}
We first prove necessity. By the definition of $U_b(t)$, we have
\begin{equation*}
    \int_{\underline{q}}^{q_\ell} [v(q, t) - P_b(t)] g(q) \,\dd q = \int_{\underline{q}}^{q_\ell} [v(q, t') - P_b(t')] g(q) \,\dd q + (t-t') R^{q_\ell}.
\end{equation*}
Substituting $v(q,t) - v(q,t') = (t-t')R^{q_\ell}$ into the above equality, we obtain
\begin{equation*}
    (t-t') R^{q_\ell}
    = \int_{\underline{q}}^{q_\ell} [P_b(t) - P_b(t')] g(q) \,\dd q + (t-t') R^{q_\ell}.
\end{equation*}
Equivalently, this implies
\begin{equation*}
    \int_{\underline{q}}^{q_\ell} [P_b(t) - P_b(t')] g(q) \,\dd q =0.
\end{equation*}
This equality holds if and only if $P_b(t)=P_b(t')$. Because $t$ and $t'$ are arbitrary, $P_b$ must be constant over $T$.

Next, we prove sufficiency. Suppose $P_b(t)=P_b(t')$ for all $t,t'\in T$. Then, according to the definition of $U_b(t)$, we have
\begin{align*}
    U_b(t) =& \int_{\underline{q}}^{q_\ell}  [v(q, t) - P_b(t)] g(q)\,\dd q\\
    =& \int_{\underline{q}}^{q_\ell}  [v(q, t) - P_b(t')] g(q)\,\dd q\\
    =& \int_{\underline{q}}^{q_\ell}  [v(q, t') - P_b(t') + v(q, t) - v(q, t')] g(q)\,\dd q\\
    =& U_b(t') + (t-t') R^{q_\ell}.
\end{align*}
Thus constraint \eqref{eq:IC property} is satisfied.
\hfill \Halmos
\end{proof}

\subsection{Proof of Lemma \ref{lem:buyer feasibility}}

\begin{proof}{Proof.}
Under $\pi^{q_\ell}$, the buyer's utility $U_b(t)$ is linear and increasing in $t$. Therefore, to ensure $U_b(t)\ge 0$ for all $t\in T$, it suffices to check the lowest type $\underline{t}$. In particular, the condition $U_b(\underline{t}) \ge 0$ requires
\begin{equation*}
    \underline{t}\cdot R^{q_\ell} - G(q_\ell) P_b \ge 0.
\end{equation*}
Rearranging the inequality yields
\begin{equation*}
    P_b \le \frac{\underline{t}\cdot R^{q_\ell}}{G(q_\ell)}
    = \underline{t}\cdot \E \left[\alpha(q)\mid q\in[\underline{q},q_\ell]\right].
\end{equation*}
Feasibility additionally requires that $U_b(t)\ge v(t)-P_b$ for all $t\in T$. This condition is equivalent to
\begin{equation*}
    \int_{\underline{q}}^{q_\ell} [v(q,t)-P_b] g(q)\,\dd q \ge \int_{\underline{q}}^{\bar{q}} [v(q,t)-P_b] g(q)\,\dd q 
\end{equation*}
Rearranging terms yields
\begin{equation*}
    \int_{q_\ell}^{\bar{q}} [v(q,t)-P_b] g(q)\,\dd q \le 0.
\end{equation*}
Solving the above inequality for $P_b$ implies
\begin{equation*}
    P_b \ge \frac{t\cdot\int_{q_\ell}^{\bar{q}} \alpha(q) g(q)\,\dd q}{1-G(q_\ell)} .
\end{equation*}
Since $\int_{q_\ell}^{\bar{q}} \alpha(q) g(q)\,\dd q>0$, the tightest constraint occurs at $t=\bar{t}$, yielding
\begin{equation*}
    P_b \ge \bar{t}\cdot \E\Big[\alpha(q)\mid q\in[q_\ell,\bar{q}]\Big].
\end{equation*}
Combining the two conditions completes the proof.
\hfill \Halmos
\end{proof}

\subsection{Proof of Theorem \ref{the:impossible}}

\begin{proof}{Proof.}
We first consider the always-trade mechanism, in which $\pi(q, t) = 1$ for all $t\in T$ and $q\in Q$. To ensure the seller's obedience, we require that
\begin{equation*}
    P_+ \ge c(q), \forall q\in Q.
\end{equation*}
Since $c(q) $ is increasing in $q$, it suffices to impose 
\begin{equation*}
    P_+ \ge c(\bar{q}).
\end{equation*}
Next, we verify the buyer's individual rationality. Under the always-trade recommendation, the buyer's expected utility is
\begin{equation*}
    U_b(t ) = v(t) - P.
\end{equation*}
Thus, individual rationality requires
\begin{equation*}
    U_b(t) \ge 0,  \forall t\in T,
\end{equation*}
which implies that 
\begin{equation*}
    P \le v(t), \forall t\in T .
\end{equation*}

Since $v(t)$ is increasing in $t$, this condition is satisfied if and only if
\begin{equation*}
    P \le v(\bar{t}).
\end{equation*}
Finally, the incentive constraint $U_b(t) \ge v(t) - P_b(t')$ is automatically satisfied, because following the recommended action coincides with any deviation that always trades.

We now turn to the never-trade mechanism, in which $\pi(q, t) = 0$, for all $t\in T$ and $q\in Q$. Seller obedience requires that 
\begin{equation*}
    P_+ \le c(q), \forall q\in Q.
\end{equation*}
Since $c(q)$ is increasing, it suffices to impose
\begin{equation*}
    P_+ \le c(\underline{q}).
\end{equation*}
Under the never-trade recommendation, the buyer's utility is identically zero: $U_b(t) = 0$. To ensure that the buyer does not deviate by trading unilaterally, we require that
\begin{equation*}
    U_b(t) \ge v(t) - P, \forall t\in T.
\end{equation*}
This condition implies
\begin{equation*}
    0 \ge v(t) - P,
\end{equation*}
which is equivalent to
\begin{equation*}
    P \ge v(t), \forall  t\in T .
\end{equation*}

Since $v(t)$ is increasing in $t$, it suffices to require
\begin{equation*}
    P \ge v(\bar{t}).
\end{equation*}
This completes the proof.
\hfill \Halmos
\end{proof}



\subsection{Proof of Lemma \ref{lem:rewrite obj}}

\begin{proof}{Proof.}
According to Equation \eqref{eq:buyer utility}, we can rewrite the first term in the objective as follows:
\begin{equation*}
	\int_T f(t) \int_Q P_b(t) \pi(q, t) g(q) \,\dd q\dd t = \int_T f(t) \left[ \int_Q v(q, t) \pi(q, t) g(q) \,\dd q - U_b(t)   \right] \, \dd t.
\end{equation*}
Based on Condition \eqref{eq:IC property}, the second term can be further expanded:
\begin{align*}
	\int_T f(t) U_b(t) \, \dd t
	=& \int_T f(t) \left[  U_b( \underline{t} ) + \int_{\underline{t}}^t R_b^{\pi}(x) \,\dd x  \right] \,\dd t \\
	=& U_b(\underline{t}) + \int_T \int_{\underline{t}}^t f(t) R_b^{\pi}(x)\,\dd x \dd t\\
	=& U_b(\underline{t}) + \int_T \int_x^{\bar{t}} f(t) R_b^{\pi}(x)\,\dd t \dd x \\
	=& U_b(\underline{t}) + \int_T [1- F(x)] R_b^{\pi}(x) \,\dd x.
\end{align*}
Then we obtain:
\begin{align*}
	&\int_T f(t) \int_Q P_b(t) \pi(q, t) g(q) \,\dd q\dd t\\
    =&- U_b(\underline{t}) + \int_Q \int_T f(t) g(q)\pi(q, t)\alpha(q) \left[ t-  \frac{1- F(t)}{f(t)}\right]\,\dd t \dd q.
\end{align*}
Putting the second term in the objective back, we obtain:
\begin{equation*}
	- U_b(\underline{t}) + \int_Q\int_T \eta(q, t) \pi(q, t)  g(q)f(t)  \,\dd t\dd q .
\end{equation*}
This completes the proof.
\hfill \Halmos
\end{proof}

\subsection{Proof of Lemma \ref{lem:eta single root}}

\begin{proof}{Proof.}
Taking the partial derivative of $\eta(q, t)$ with respect to $q$ and $t$, we obtain:
\begin{align}
    &\eta_q = \frac{\partial \eta(q, t)}{\partial q } = \alpha'(q) \left[ t - \frac{1- F(t)}{f(t)} \right]  ,\label{eq:eta_q} \\
    &\eta_t = \frac{\partial \eta(q, t)}{\partial t} = \alpha(q) \left[1 - \left( \frac{1-F(t)}{f(t)} \right)' \right ] > 0 .\label{eq:eta_t}
\end{align}
If $t - \frac{1- F(t)}{f(t)} \le 0$, then $\eta(q, t), \forall q \in Q$, meaning that the function is everywhere non-positive. Otherwise, $\eta(q, t)$ is strictly increasing in $q$. Therefore, given any $t$, there exists at most one quality threshold $q_0$ such that $\eta(q_0, t) = 0 $.
\hfill \Halmos
\end{proof}

\subsection{Proof of Theorem \ref{prop:relax optimal}}

\begin{proof}{Proof.}
It suffices to show that mechanism $(\pi^*, P_b^*)$ maximizes the revenue equation \eqref{eq:rewrite obj} and satisfies all constraints in Lemma \ref{lem:IC property}.

We first verify optimality. Notice that the second term in Equation \eqref{eq:rewrite obj} can be maximized pointwise in $(q,t)$: sets $\pi(q,t) = 1$ whenever $\eta(q,t) \ge 0$ and $\pi(q,t) = 0$ otherwise. By the definition of $\lambda(t)$, it is straightforward to check that $\pi^*$ is exactly constructed this way. Therefore, $\pi^*$ pointwise maximizes the second term in Equation \eqref{eq:rewrite obj}.

By Condition \eqref{eq:IR property}, we have $U_b(\underline{t}) \ge 0$. Hence, the first term $-U_b(\underline{t})$ in Equation \eqref{eq:rewrite obj} is upper bounded by 0. Letting $U_b(\underline{t}) = 0$ achieves this upper bound, which yields the optimal payment function for the buyer:
\begin{align}
\label{eq:payment}
	P_b^*(t) &= \frac{\int_Q \pi^*(q, t) v(q, t) g(q) \,\dd q  - \int_{\underline{t}}^t R^{\pi^*}_b(x)\,\dd x }{\int_Q \pi^*(q, t) g(q)\,\dd q} \nonumber \\
	&= \E [v(q, t) \mid q\ge \lambda(t)] - \frac{\int_{\underline{t}}^t R^{\pi^*}_b(x)\,\dd x}{1- G(\lambda(t))}.
\end{align}
Note that the above equation only applies to the set of types satisfying $1- G(\lambda(t)) > 0$. By the definition of $t_1$, this set is exactly the interval $[t_1, \bar{t}]$.
\begin{itemize}
	\item If $t_1 \le \underline{t}$, Equation \eqref{eq:payment} holds for all $t\in T$.
	\item If $t_1 > \underline{t}$, Equation \eqref{eq:payment} applies only to $t \in [t_1, \bar{t}]$. For buyers with type $t \in [\underline{t}, t_1)$, who are never recommended to trade, the payment should be set arbitrarily high. Without loss of generality, we set $P_b^*(t) = v(\bar{q}, t_1)$.
\end{itemize}

Next, we show that $\pi^*$ satisfies the monotonicity constraint \eqref{eq:monotone}. Plugging $\pi^*$ into $R_b^{\pi}(t)$, we get:
\begin{equation}
\label{eq:optimal R}
	R_b^{\pi^*}(t) = \int_{\lambda(t)}^{\bar{q}} \alpha(q) g(q) \,\dd q.
\end{equation}
Based on the implicit differentiation method, we have:
\begin{equation}
\label{eq:lambda monotone}
	\frac{\dd \lambda(t)}{\dd t} = -\frac{\eta_t}{\eta_q } < 0, \forall t\in T.
\end{equation}
This inequality is due to Equation \eqref{eq:eta_q} and Equation \eqref{eq:eta_t}. Combined with $\alpha(q) >0$ for all $q$, we obtain $R_b^{\pi^*}(t)$ is monotone increasing in $t$.

In summary, $(\pi^*, P_b^* )$ indeed is the optimal solution to the relaxed problem \eqref{eq:relaxed LP}.
\hfill \Halmos
\end{proof}

\subsection{Proof of Lemma \ref{prop:reverse}}

\begin{proof}{Proof.}
We prove in two steps. First, we establish a technical claim. Then we complete the proof of the lemma using that claim.

\paragraph{Claim.} Let $x$ be a random variable with CDF $G$ and PDF $g$. Let $e(y) = \E [\alpha(x) \mid x \ge y ]$. Then, 
\begin{equation*}
	e'(y) = \frac{ g(y) \int_{y}^{\bar{x}} [1- G(x)] \alpha'(x) \,\dd x }{ [1- G(y) ]^2  }.
\end{equation*} 

\begin{proof}{Proof of the Claim.}
Note that
\begin{align*}
	e(y) =& \frac{\int_y^{\bar{x}} \alpha(x) g(x) \,\dd x }{1 - G(y) } \\
    =& \alpha(y) + \frac{\int_y^{\bar{x}} [1- G(x)] \alpha'(x)\,\dd x }{1 - G(y) },
\end{align*}
where the first equality follows from the definition and the second one comes from integration by parts. Then
\begin{align*}
	e'(y) =& \alpha'(y) - \frac{\alpha'(y) [1-G(y)]^2 - g(y) \int_y^{\bar{x}} \alpha'(x) [1- G(x)] \,\dd x  }{[1-G(y)]^2 }\\
	=& \frac{  g(y) \int_y^{\bar{x}} \alpha'(x) [1- G(x)] \,\dd x}{[1-G(y)]^2 }. \tag*{\Halmos}
\end{align*}
\end{proof}

With the Claim in hand, we proceed to show the properties of $P_b^*$. According to Equation \eqref{eq:optimal buyer pay}, $P_b^*$ is a constant in $[\underline{t}, t_1)$. We first show that $P_b^*$ is also a constant in  $[t_2, \bar{t}]$.

If $t \in [t_2, \bar{t}] $, the buyer will be recommended to buy with probability 1, and $P_b^*$ becomes:
\begin{align*}
	P_b^*(t) =& \E_q[\alpha(q)] \cdot t  - ( t- t_1 )\cdot \E_q[\alpha(q)] \\
    =& \E_q[\alpha(q)] t_1.
\end{align*}
Obviously, $P_b^*$ is  continuous at $t_2$. Next, we show that it is continuous at $t_1$ when $t_1 > \underline{t}$. 
\begin{align*}
	\lim \limits_{t \rightarrow t_1^+} P_b^*(t) =& \lim \limits_{t \rightarrow t_1^+} \left[ \frac{\int_Q v(q, t) \pi^*(q, t) g(q)\,\dd q - \int_{t_1}^t R_b^{\pi^*}(x)\, \dd x }{1- G(\lambda(t)) } \right] \\
	=& \lim \limits_{t \rightarrow t_1^+} \left[ \frac{\int_{\lambda(t)}^{\bar{q}} v(q, t) g(q)\,\dd q - \int_{t_1}^t R_b^{\pi^*}(x)\, \dd x }{1- G(\lambda(t)) } \right] \\
	=& \lim \limits_{t \rightarrow t_1^+} \left[ \frac{R_b^{\pi^*}(t) -v(\lambda(t), t) g(\lambda(t)) \lambda'(t) - R_b^{\pi^*}(t)}{-g(\lambda(t)) \lambda'(t)  } \right]\\
	=& v(\bar{q} , t_1),
\end{align*}
where the second line is the definition of $P_b^*$, and the third line follows from the structure of $\pi^*$. Using the L'Hospital rule, we can obtain the fourth line.

Then we show that $P_b^*$ is strictly decreasing in $[t_1, t_2]$. For $P_b^*(t)$, we take derivative with respect to $t$:
\begin{align*}
	\frac{\dd P_b^*(t)}{\dd t} =& \frac{\dd}{\dd t} \left[ t\cdot \E[\alpha(q) \mid q\ge \lambda(t)] - \frac{\int_{t_1}^t R_b^{\pi^*} (x) \,\dd x }{1- G(\lambda(t))} \right] \\
	=& \frac{\int_{\lambda(t)}^{\bar{q}} \alpha(q) g(q)\,\dd q }{1-G(\lambda(t))} + t\cdot \frac{g(\lambda(t))\int_{\lambda(t)}^{\bar{q}}[1-G(q)] \alpha'(q)\,\dd q  }{[1-G(\lambda(t))]^2} \cdot \lambda'(t) \\
	&- \left[ \frac{R_b^{\pi^*}(t) }{ 1- G(\lambda(t))} + \frac{g(\lambda(t)) \int_{t_1}^t R_b^{\pi^*}(x)\,\dd x  }{[1- G(\lambda(t))]^2} \cdot \lambda'(t)  \right]\\
	=& -\frac{g(\lambda(t)) \lambda'(t) }{[1 - G(\lambda(t))]^2} \left[ \int_{t_1}^t R_b^{\pi^*}(x)\,\dd x - \int_{\lambda(t)}^{\bar{q}} [1- G(q)] \alpha'(q) t \,\dd q  \right],
\end{align*}
where the second line follows from the definition of $P_b^*(t)$ and the third line is due to the claim. The last equation comes from $R_b^{\pi^*}(t) = \int_{\lambda(t)}^{\bar{q}} \alpha(q)g(q) \,\dd q$.

In order to show $\frac{\dd P_b^*(t)}{\dd t} <0$, it suffice to show 
\begin{equation*}
	\int_{t_1}^t R_b^{\pi^*}(x)\,\dd x - \int_{\lambda(t)}^{\bar{q}} [1- G(q)] \alpha'(q) t \,\dd q < 0,
\end{equation*}
since $\lambda'(t) < 0$. Note that the above inequality is true when $t= t_1$, since $\lambda(t_1) \le \bar{q}$. Next, we show that the above equation is monotone decreasing in $[t_1, t_2]$:
\begin{align*}
	&\frac{\dd}{\dd t} \left[ \int_{t_1}^t R_b^{\pi^*}(x)\,\dd x - \int_{\lambda(t)}^{\bar{q}} [1- G(q)] \alpha'(q) t \,\dd q  \right]\\
	=& R_b^{\pi^*} (t) - \left[\int_{\lambda(t)}^{\bar{q}} [1- G(q)] \alpha'(q) \,\dd q - [1 - G(\lambda(t))] \alpha'(\lambda(t)) t \lambda'(t)   \right]\\
	=& [ 1- G(\lambda(t))  ] [\alpha(\lambda(t)) + \alpha'(\lambda(t)) \cdot t \cdot \lambda'(t) ]\\
	=& [ 1- G(\lambda(t))  ] \left[\alpha(\lambda(t)) - \frac{ \alpha'(\lambda(t))\cdot t \cdot \eta_t }{ \eta_q } \right] \\
	= & [ 1- G(\lambda(t))  ] \frac{\alpha(\lambda(t)) \left[ t\cdot \left(\frac{1-F(t)}{f(t)} \right)' - \frac{1-F(t)}{f(t)} \right] }{ \left[t - \frac{1 - F(t)}{f(t)} \right] }<0,
\end{align*}
where the third line follows from integrating $R_b^{\pi^*}$ by parts, the fourth line follows from Equation \eqref{eq:lambda monotone}. The fifth line comes from Equation \eqref{eq:eta_q} and \eqref{eq:eta_t}. The last inequality uses the MHR condition. Therefore, we obtain:
\begin{equation*}
	\int_{t_1}^t R_b^{\pi^*}(x)\,\dd x - \int_{\lambda(t)}^{\bar{q}} [1- G(q)] \alpha'(q) t \,\dd q < 0.
\end{equation*}
This completes the proof.
\hfill \Halmos
\end{proof}

\subsection{Proof of Theorem \ref{prop:origin optimal}}

\begin{proof}{Proof.}
We need to verify that $U_b(t) \ge  \max_{t'} v(t) - P_b^*(t')$. Before that, we need to show that $\E[v(q, t) \mid q < \lambda(t)] - P_b^*(t) \le 0 $ for all $t$. That is, if recommended not to buy, the buyer will follow the recommendation. Note that for $t \in [t_1, \bar{t}]$,
\begin{align}
\label{eq:tools for theorem 2}
	\E[v(q, t) \mid q < \lambda(t) ] - P_b^*(t) \le & v(\lambda(t), t) - P_b^*(t) \nonumber \\
	= & v(\lambda(t), t) - \left[ \E [v(q, t) \mid q\ge \lambda(t)] - \frac{\int_{t_1}^t R_b(x)\,\dd x}{1- G(\lambda(t))} \right] \nonumber \\
	=& \frac{1}{ 1- G(\lambda(t))} \left(\int_{t_1}^t R_b^{\pi^*}(x)\,\dd x - \int_{\lambda(t)}^{\bar{q}} [1-G(q)] \alpha'(q) t \,\dd q  \right)  \nonumber \\
	<& 0,
\end{align} 
where the last inequality follows from the analysis in Lemma \ref{prop:reverse}. Obviously, for buyers in $ [\underline{t}, t_1)$, they will not want to buy. Therefore, the buyer will follow the ``no trade'' recommendation.

According to Lemma \ref{prop:reverse}, we have $\min_t P_b^*(t) = P_b^*(\bar{t})$. So for any $t\in [\underline{t}, \bar{t} ]$, we have:
\begin{align*}
	\max_{t'}\left\{ v(t) - P_b^*(t')\right\}  =& t \cdot \E_q[\alpha(q)] - P_b^*(\bar{t})  \\
	= & [1- G(\lambda(\bar{t}))] [ t\cdot \E_q [\alpha(q) \mid q\ge \lambda(\bar{t}) ] - P_b^*(\bar{t}) ]  \\
	&+  G(\lambda(\bar{t})) [ t \cdot \E_q[\alpha(q) \mid q < \lambda(\bar{t}) ] - P_b^*(\bar{t}) ]\\
	\le&[1- G(\lambda(\bar{t}))] [ t\cdot \E_q [\alpha(q) \mid q\ge \lambda(\bar{t}) ] - P_b^*(\bar{t}) ] \\
	&+  G(\lambda(\bar{t})) [ \bar{t} \cdot \E_q[\alpha(q) \mid q < \lambda(\bar{t}) ] - P_b^*(\bar{t}) ]\\
	\le& [1- G(\lambda(\bar{t}))] [ t\cdot \E_q [\alpha(q) \mid q\ge \lambda(\bar{t}) ] - P_b^*(\bar{t}) ] \\
	=& U_b(\bar{t}; t),
\end{align*}
where the first inequality follows from $t \le \bar{t}$, and the second comes from \eqref{eq:tools for theorem 2}. Therefore, we obtain
\begin{equation*}
	U_b(t) \ge \max_{t'} \left\{v(t) - P_b^*(t')\right\},
\end{equation*}
since we have constraint $U_b(t) \ge U_b(t'; t), \forall t, t'\in T$.
\hfill \Halmos
\end{proof}

\subsection{Proof of Lemma \ref{lem:partial IC property}}

\begin{proof}{Proof.}
We first prove that the constraints are necessary for the buyer. Slightly manipulating the buyer's IC constraint \eqref{eq:partial IC buyer} yields:
\begin{align*}
    U_b(t) &\ge U_b(t') + \int_{q\in Q} \pi(q, t')[v(q,t) - v(q, t')] g(q) \,\dd q \\
    &= U_b(t') + (t-t')\int_{q\in Q} \alpha(q) \pi(q, t') g(q) \,\dd q.
\end{align*}

Note that $R_b^{\pi}(t) = \int_{q \in Q} \alpha(q) \pi(q, t)g(q)  \,\dd q$, thus the above inequality is equivalent to:
\begin{equation}
\label{eq:buyer_ic_1}
    U_b(t) - U_b(t') \ge (t-t')R^{\pi}_b(t').
\end{equation}
The above equation should hold for any $t$ and $t'$. As a result, if we switch $t$ and $t'$, we should have:
\begin{equation}
\label{eq:buyer_ic_2}
    U_b(t') - U_b(t) \ge (t'-t)R^{\pi}_b(t).
\end{equation}
Combining Equation \eqref{eq:buyer_ic_1} and \eqref{eq:buyer_ic_2} gives:
\begin{equation}
    (t-t') R^{\pi}_b(t') \le U_b(t) - U_b(t') \le (t-t')R^{\pi}_b(t).
\end{equation}
When $t>t'$, we can divide the above inequality by $t-t'$ and get:
\begin{equation}
    R^{\pi}_b(t') \le \frac{U_b(t) - U_b(t')}{t - t'} \le R^{\pi}_b(t),
\end{equation}
Letting $t \rightarrow t'$, the above inequalities become:
\begin{equation}
    \frac{\mathrm{d} U_b(t)}{\mathrm{d}t} = R^{\pi}_b(t).
    \label{eq:utility_derivative}
\end{equation}
The above equation still holds if $t<t'$. Therefore, Equation \eqref{eq:partial IC property buyer} follows. Adding Equation \eqref{eq:buyer_ic_1} and \eqref{eq:buyer_ic_2} gives $(t-t')[R_b^{\pi}(t)-R_b^{\pi}(t')]\ge 0$,
which implies constraint \eqref{eq:partial buyer_monotone}.

Now, let's consider the IR constraint for the buyer. By definition, $R^{\pi}_b(t)$ is non-negative. Thus, with Equation \eqref{eq:partial IC property buyer}, we know that $U_b(t)$ attains its minimum value at $\underline{t}$. Therefore, to satisfy the IR constraint, we only need to ensure $U_b(\underline{t})\ge 0$, which is exactly Equation \eqref{eq:partial IR property buyer}.

We omit the proof for constraint \eqref{eq:partial seller_monotone}, Equation \eqref{eq:partial IC property seller}, and Equation \eqref{eq:partial IR property seller}, as they can be obtained through similar analysis for the seller.

Now we show that the constraints are also sufficient for the buyer. The IC constraint \eqref{eq:partial IC buyer} for the buyer is equivalent to:
\begin{equation*}
    U_b(t) \ge U_b(t') + (t-t')\int_{q \in Q} \alpha(q) \pi(q, t')g(q) \,\dd q.
\end{equation*}
Constraint \eqref{eq:partial IC property buyer} implies the above inequality, because if $t' < t$, we have
\begin{equation*}
    U_b(t)-U_b(t') = \int_{t'}^t R^{\pi}_b(x)\,\mathrm{d}x \ge (t -t') R^{\pi}_b(t').
\end{equation*}
Similarly, when $t' > t$, we also have $U_b(t) - U_b(t') \ge  (t -t') R^{\pi}_b(t') $.

The IR constraint is equivalent to $U_b(t) \ge 0$. $R^{\pi}_b(t') \ge 0$ and $R^{\pi}_b(t)$ is monotone non-decreasing implies $R^{\pi}_b(t) \ge 0, \forall t$. Therefore, Equation \eqref{eq:partial IC property buyer} and \eqref{eq:partial IR property buyer} together show that $U_b(t) \ge 0$ for any $t$.
\hfill \Halmos
\end{proof}

\subsection{Proof of Lemma \ref{lem:partial equal revenue}}

\begin{proof}{Proof.}
For any feasible mechanism $(\pi, P_b, P_s)$, the revenue of the mediator can be written as:
\begin{align}
\label{eq:revenue}
\begin{aligned}
    \int_{t} f(t) \int_{q}g(q)\pi(q, t) P_b(t) \,\dd q\dd t -\int_{q}g(q) \int_{t } f(t) \pi(q, t) P_s(q) \,\mathrm{d}t\mathrm{d}q .
\end{aligned}
\end{align}

Using Equation \eqref{eq:buyer utility}, we have:
\begin{align*}
    &\int_{t} f(t) \int_{q}g(q)\pi(q, t) P_b(t)\,\dd q\dd t \\
    =& \int_{t} f(t) \left[\int_{q}\pi(q, t) v(q, t) g(q)\,\dd q - U_b(t)  \right] \,\dd t, 
\end{align*}
where the second term containing $U_b(t)$ can be further expanded as follows:
\begin{align*}
    \int_{t} f(t) U_b(t) \,\dd t
    =& \int_{t} f(t) \left[ \int_{\underline{t}}^t R^{\pi}_b(x)\,\dd x + U_b(\underline{t})\right] \,\dd t\\
    =&\int_{t}\int_{\underline{t}}^t f(t)R^{\pi}_b(x)\,\dd x\dd t + U_b(\underline{t})\\
    =& \int_{t}\int_{x}^{\bar{t}} f(t)R^{\pi}_b(x)\,\dd t\dd x + U_b(\underline{t})\\
    = &\int_{t}[1-F(x)]R^{\pi}_b(x)\,\dd x + U_b(\underline{t})\\
    =& \int_{q} g(q)\left[\int_{t}[1-F(t)]\alpha(q) \pi(q, t) \,\dd t \right]\,\dd q + U_b(\underline{t}).
\end{align*}
Plugging the term back into the above equation, and switching the order of the integral gives:
\begin{align*}
    &\int_{t} f(t) \int_{q}g(q)\pi(q, t) P_b(t)\,\dd q\dd t\\
    =&\int_{q} g(q) \left[\int_{t} f(t)\pi(q, t)\alpha(q) \phi^-_b(t)  \,\dd t \right]\,\dd q - U_b(\underline{t}).
\end{align*}

The second term of Equation \eqref{eq:revenue} is about the seller. According to Equation \eqref{eq:seller utility}, it can be written as:
\begin{align}
\label{eq:rev_seller}
    &\int_{q} g(q) \int_{t} f(t) \pi(q, t)P_s(q) \,\dd t\dd q \nonumber\\
    =&\int_{q} g(q) \left[\int_{t}\pi(q, t)c(q)f(t)\,\dd t + U_s(q) \right] \,\dd q,
\end{align}
where we can rewrite the second term as:
\begin{align*}
    \int_{q} g(q) U_s(q) \,\dd q
    =& \int_{q} g(q) \left[k\int_q^{\bar{q}} R^{\pi}_s(x)\,\dd x + U_s(\bar{q}) \right]\,\dd q\\
    =&k\int_{q} g(q)\int_{q}^{\bar{q}}R^{\pi}_s(x)\,\dd x\dd q + U_s(\bar{q})\\
    =&k\int_{q}\int_{\underline{q}}^{x}g(q)R^{\pi}_s(x)\,\dd q\dd x + U_s(\bar{q})\\
    =&k\int_{q}G(x)R^{\pi}_s(x)\,\dd x + U_s(\bar{q})\\
    =&k\int_{t}f(t)\int_{q}G(q)\pi(q, t)\,\dd q \dd t + U_s(\bar{q})
\end{align*}
Putting everything back to Equation \eqref{eq:rev_seller} yields:
\begin{align*}
    &\int_{q} g(q) \int_{t} f(t) \pi(q, t)P_s(q) \,\dd t\dd q \\
    = &\int_{t } f(t) \left[\int_{q}g(q)\pi(q, t) k\left(q +\frac{G(q)}{g(q)}\right)\,\dd q\right]\,\dd t + U_s(\bar{q})\\
    =&\int_{t} f(t) \left[\int_{q}g(q)\pi(q, t) k\phi^+_s(q)\,\dd q\right]\,\dd t + U_s(\bar{q}).
\end{align*}
By combining both two terms of Equation \eqref{eq:revenue}, we obtain:
\begin{equation*}
    -U_b(\underline{t})-U_s(\bar{q}) + \int_{q} \int_{t}\pi(q,t)\left[\alpha(q) \phi^-_b(t)  - k \phi^+_s(q)\right]f(t)g(q) \,\dd t\dd q.
\end{equation*}
This completes the proof.
\hfill \Halmos
\end{proof}

\subsection{Proof of Theorem \ref{thm:optimal_mechanism}}

\begin{proof}{Proof.}
We first show that the mechanism $(\pi^*, P_b^*, P_s^*)$ defined in Proposition \ref{thm:optimal_mechanism} is feasible. 

Observe that if a problem instance is regular, the signaling rule $\pi^*(q, t)$ is non-decreasing in $t$ but non-increasing in $q$. This implies that $R^{\pi^*}_b(t)$ is non-decreasing in $t$ and $R^{\pi^*}_s(q)$ is non-increasing in $q$. Satisfying constraint \eqref{eq:partial buyer_monotone} and \eqref{eq:partial seller_monotone}.

By definition, the buyer's utility is:
\begin{align*}
    U_b(t) =& \int_{q} \pi^*(q, t)[v(q, t) - P^*_b(t)] g(q)\,\dd q \\
    =&\int_{\underline{t}}^t \int_{q}\alpha(q)\pi^*(q, x)g(q) \,\dd q \dd x.
\end{align*}
This implies that $U_b(\underline{t})=0$ and 
\begin{align*}
	\frac{\dd U_b(t)}{\dd t}=&\int_{q}\alpha(q)\pi^*(q, t)g(q) \,\dd q\\
    =&R_b^{\pi^*}(t).
\end{align*}
Therefore, Condition \eqref{eq:partial IC property buyer} follows. Similarly, by definition, the seller's utility is:
\begin{align*}
    U_s(q) =& \int_{t} \pi^*(q, t)[P_s^*(q) - c(q)]f(t)\,\dd t \\
    =& k\int_{q}^{\bar{q}} \int_{t}\pi^*(x, t)f(t) \,\dd t \dd x.
\end{align*}
It follows that $U_s(\bar{q})=0$ and
\begin{align*}
	\frac{\dd U_s(q)}{\dd q}=&-k\int_{t}\pi^*(q, t)f(t) \,\dd t\\
    =&-kR_s^{\pi^*}(q),
\end{align*}
which clearly implies Condition \eqref{eq:partial IC property seller}. Therefore, $(\pi^*, P_b^*, P_s^*)$ is a feasible mechanism.

Then we prove the optimality. The revenue equation \eqref{eq:partial equal revenue} contains 3 terms. We now show that the mechanism $(\pi^*, P_b^*, P_s^*)$ optimizes them all at the same time. 

By definition, $\pi^*(q, t)=1$ if and only if $\phi_b^-(t)\ge \frac{k\phi_s^+(q)}{\alpha(q)}$, which is equivalent to $\alpha(q) \phi^-_b(t)- k \phi^+_s(q)\ge 0$. This means that the last term is point-wise optimized for all $(t,q)$ pairs, and thus the term is maximized.

For the first and second terms, if we want to maximize the revenue, we need to minimize both these terms. According to Lemma \ref{lem:partial IC property}, both $U_b(\underline{t})$ and $U_s(\bar{q})$ need to be non-negative. The proof of feasibility already shows that $U_b(\underline{t})=0$ and $U_s(\bar{q})=0$ under mechanism $(\pi^*, P_b^*, P_s^*)$. Thus, both terms are optimized. 
\hfill \Halmos
\end{proof}

\subsection{Proof of Proposition \ref{prop:decreasing}}


\begin{proof}{Proof.}
\begin{align*}
    U_b(t) - v(t)
    =& \Big[ U_b(\underline{t})  + \int_{\underline{t}}^{t}\int_q \pi(q, x) \alpha(q) g(q) \,\dd q \dd x \Big] - \int_q \alpha(q) t g(q) \, \dd q\\
    =& \Big[ U_b(\underline{t})  + \int_{\underline{t}}^{t}\int_q \pi(q, x) \alpha(q) g(q) \,\dd q \dd x \Big] - \Big[ \int_{\underline{t}}^t \int_q \alpha(q) g(q) \,\dd q \dd x + v(\underline{t}) \Big]\\
    =& U_b(\underline{t}) - v(\underline{t}) + \Big[\int_{\underline{t}}^{t}\int_q [\pi(q, x)-1] \alpha(q) g(q) \,\dd q \dd x \Big] .
\end{align*}
Since $\pi(q, x) - 1 \le 0$ and $\alpha(q) g(q) \ge 0$, we have that $U_b(t) - v(t)$ is monotone decreasing in $t$.
\hfill \Halmos
\end{proof}

\subsection{Proof of Theorem \ref{thm:optimal_mechanism_irregular}}

\begin{proof}{Proof.}
We begin by applying the classical ironing method to $\psi(t)$. By definition, the mediator's revenue function can be rewritten as
\begin{align*}
	&\int_{\underline{q}}^{\bar{q}} \int_{\underline{t}}^{\bar{t}} \pi(q, t)\alpha(q)[\psi(t) - \varphi(q) ]  \,\dd F(t)\dd G(q)\\
	=& \int_{\underline{q}}^{\bar{q}} \int_{\underline{t}}^{\bar{t}} \pi(q, t)\alpha(q)[\bar{\psi}(t) - \varphi(q) ]  \,\dd F(t)\dd G(q)\\
	&+\int_{\underline{q}}^{\bar{q}} \int_{\underline{t}}^{\bar{t}}\pi(q, t)\alpha(q)[h_b(F(t)) \\
    & - l_b(F(t))] \,\dd F(t)\dd G(q),
\end{align*}
since $h_b(F(t)) = \psi(t), l_b(F(t)) = \bar{\psi}(t)$.
We can simplify the second term above via integration by parts:
\begin{align*}
	&\int_{q} \int_{t} \pi(q, t)\alpha(q)[h_b(F(t)) - l_b(F(t))] \, \dd F(t) \dd G(q)\\
	=& \int_{\underline{t}}^{\bar{t}} [h_b(F(t)) - l_b(F(t))] R_b^{\pi}(t) \,\dd F(t)\\
	=&[H_b(F(t)) - L_b(F(t))] R_b^{\pi}(t)|_{\underline{t}}^{\bar{t}}- \int_{\underline{t}}^{\bar{t}} [H_b(F(t)) - L_b(F(t))] \,\dd R_b^{\pi}(t).
\end{align*}

By definition, $L_b$ is the convex envelope of $H_b$, then we have $L_b(0)=H_b(0)$ and $L_b(1) = H_b(1)$. Therefore, the first term is 0, and the revenue function becomes:
\begin{equation}
\label{eq:iron buyer}
	\int_{\underline{q}}^{\bar{q}} \int_{\underline{t}}^{\bar{t}}  \pi(q, t)\alpha(q)[\bar{\psi}(t) - \varphi(q) ]\,\dd F(t)\dd G(q) -\int_{\underline{t}}^{\bar{t}} [H_b(F(t)) - L_b(F(t))] \,\dd R_b^{\pi}(t).
\end{equation}

\textbf{Tailored Ironing.} To iron a non-monotone function $\varphi(q)$ with a scaling term $\alpha(q)$, we propose a variant ironing approach tailored to our setting. Specifically, we modify the standard ironing by changing the definition of $w = G(q)$ to the following weighted cumulative distribution function:
\begin{equation*}
    w(q) = \int_{\underline{q}}^q\alpha(r)g(r) \,\dd r.
\end{equation*}
Note that $\alpha(q)>0$ and $g(q) > 0$ for all $q\in [\underline{q}, \bar{q}]$. Thus, $w(q)$ is a strictly increasing function, and has an inverse function $w^{-1}:[0, \bar{w}] \mapsto [\underline{q}, \bar{q}]$, where $\bar{w} = \int_{\underline{q}}^{\bar{q}}\alpha(r)g(r) \,\dd r$. The remaining ironing procedure stays unchanged, and we also define the corresponding four functions as $h_s(w)$, $H_s(w)$, $l_s(w)$ and $L_s(w)$. And the ironed function $\bar{\varphi}(q) = l_s(w(q)) $.

Now, we are ready to iron function $\varphi(q)$. Similar to the case of function $\psi(t)$, the first term in Equation \eqref{eq:iron buyer} can be written as:
\begin{align*}
	&\int_{\underline{q}}^{\bar{q}} \int_{\underline{t}}^{\bar{t}} \pi(q, t)\alpha(q)[\bar{\psi}(t) - \varphi(q) ]\,\dd F(t)\dd G(q)\\
	=&\int_{\underline{q}}^{\bar{q}} \int_{\underline{t}}^{\bar{t}} \pi(q, t)\alpha(q)[\bar{\psi}(t) - \bar{\varphi}(q) ]\,\dd F(t)\dd G(q)\\
	&+\int_{\underline{q}}^{\bar{q}} \int_{\underline{t}}^{\bar{t}} \pi(q, t)\alpha(q)[l_s(w(q)) \\
    & - h_s(w(q)) ]\,\dd F(t)\dd G(q)
\end{align*}
since $h_s(w(q)) = \varphi(q)$ and $l_s(w(q)) = \bar{\varphi}(q)$. We can simplify the second term as follows:
\begin{align*}
	&\int_{\underline{q}}^{\bar{q}} \int_{\underline{t}}^{\bar{t}} \pi(q, t)\alpha(q)[l_s(w(q))- h_s(w(q)) ]\,\dd F(t)\dd G(q)\\
	=&\int_{\underline{q}}^{\bar{q}} \int_{\underline{t}}^{\bar{t}} \pi(q, t)f(t)[l_s(w(q)) - h_s(w(q)) ]\,\dd t\dd w(q)\\
	=&\int_{\underline{q}}^{\bar{q}} [l_s(w(q)) - h_s(w(q))] R_s^{\pi}(q) \,\dd w(q) \\
	=&[L_s(w(q)) - H_s(w(q))] R_s^{\pi}(q)|_{\underline{q}}^{\bar{q}} - \int_{\underline{q}}^{\bar{q}} [L_s(w(q)) - H_s(w(q))] \,\dd R_s^{\pi}(q)
\end{align*}
$L_s(0) = H_s(0)$ and $L_s(\bar{w}) = H_s(\bar{w})$ since $L_s$ is the convex envelope of $H_s$. Thus, the first term is equal to 0. Overall, the mediator's revenue can be written as
\begin{align}
\label{eq:iron rev}
	 & \int_{\underline{q}}^{\bar{q}} \int_{\underline{t}}^{\bar{t}} \pi(q, t)\alpha(q)[\bar{\psi}(t) - \bar{\varphi}(q) ]\,\dd F(t)\dd G(q) \nonumber \\
	&-\int_{\underline{t}}^{\bar{t}} [H_b(F(t)) - L_b(F(t))] \,\dd R_b^{\pi}(t) \nonumber  \\
	&-\int_{\underline{q}}^{\bar{q}} [L_s(W(q)) - H_s(W(q))] \,\dd R_s^{\pi}(q) \nonumber  \\
    &-U_b(\underline{t}) - U_s(\bar{q}) .
\end{align}


Next, we show that the mechanism $(\pi^*, P_b^*,P_s^*)$ maximizes all the terms in Equation \eqref{eq:iron rev} simultaneously. The first term is maximized by $\pi^*$, as $\pi^*(q, t) = 1$ only when $\bar{\psi}(t) - \bar{\varphi}(q) \ge 0$. Moreover, $(\pi^*, P_b^*, P_s^*)$ satisfies $U_b(\underline{t}) = 0$ and $U_s(\bar{q}) = 0$, indicating that it also maximizes the last two terms since Lemma \ref{lem:partial IC property} implies that $U_b(t_1)\ge 0$ and $U_s(q_2) \ge 0$ hold for all feasible mechanisms. 

As for the second term, it is worth noting that $H_b(F(t)) - L_b(F(t)) \ge 0$ since $L_b(w)$ is the convex envelope of $H_b(w)$. Additionally, based on Lemma \ref{lem:partial IC property}, we know that $\dd R_b^{\pi}(t) \ge 0$, which means this term is always non-negative. Therefore, in order to show that the second term is maximized, it suffices to prove that this term equals to 0. Actually, it is only necessary to consider cases where $H_b(F(t)) - L_b(F(t)) > 0$. In such cases, $t$ must fall within an ironed interval $I$, thus function $L_b(w)$ is linear in the interval $I$. This implies $l_b(w) = \bar{\psi}(t)$ is a constant and thus $R_b^{\pi}(t) = \int_{q: \bar{\varphi}(q)\le \bar{\psi}(t)} \alpha(q) g(q) \,\mathrm{d}q$ is also a constant in the interval $I$, leading to $\mathrm{d}R_b^{\pi^*}(t) = 0$.

Similarly, for the third term, we observe that $L_s(w(q)) - H_s(w(q)) \le 0$ since $L_s(w)$ is a convex envelope of $H_s(w)$. According to Lemma \ref{lem:partial IC property}, $\dd R_s^{\pi}(q) \le 0$ holds for all feasible mechanisms. Consequently, this term is also non-negative. We show that in the optima mechanism $(\pi^*,P_b^*, P_s^*)$, this term equals to 0. Consider cases where $L_s(w(q)) - H_s(w(q)) <0$. In such cases, $q$ must lie in an ironed interval $I$ and thus $L_s(w)$ is linear in the ironing interval. This implies $l_s(w) = \bar{\varphi}(q)$ is a constant. Then $R_s^{\pi^*}(q) = \int_{t:\bar{\psi}(t) \ge \bar{\varphi}(q)} f(t) \,\dd t$ is also a constant in the interval $I$, leading to $\dd R_s^{\pi^*}(q) = 0$.

In conclusion, the mechanism $(\pi^*, P_b^*, P_s^*)$ optimizes all terms in the Equation \eqref{eq:iron rev} simultaneously, proving it to be an optimal feasible mechanism.
\hfill \Halmos
\end{proof}

\end{APPENDICES}

\end{document}